\documentclass[	pra,
		twocolumn,
		groupedaddress,
		nofootinbib,
		showpacs,
		]{revtex4}
\usepackage[long,short,remarks,todo]{optional}

\usepackage{amsmath,amstext,amsfonts,amssymb,amsthm}
\usepackage[caption=false]{subfig} 
\usepackage{dsfont}					
\usepackage{upgreek}				
\usepackage{siunitx}
\usepackage[dvipsnames]{xcolor}
\usepackage{pgf}
\usepackage{tikz}
\usetikzlibrary{arrows,automata}
\newcommand{\ket}[1]{\left| #1\right>}
\newcommand{\bra}[1]{\left< #1\right|}

\newcommand{\vek}[1]{\mathbf #1}
\newcommand{\ketbra}[2]{|#1\rangle\!\langle #2|}

\newcommand{\matrixElement}[3]{\left\langle   #1 \vphantom{#3}  \right|   #2  \left| \vphantom{#1} #3 \right\rangle}
\newcommand{\matrixElementSub}[5]{\mspace{3.0mu}{}_{#2}\mspace{-5.0mu}\matrixElement{#1}{#3}{#4}\mspace{-5.0mu}{}_{#5}}
\newcommand{\ketbraIX}[3]{\ket{#1}_{\! #2}\!\!\bra{#3}}
\newcommand{\braketIX}[4]{{\vphantom \langle}_{#1}\!\left\langle #2 \!\right.\left| #3\right\rangle_{#4}}

\newcommand{\modulus}[1]{\left\lvert #1 \right\rvert}
\newcommand{\mean}[1]{\left< #1 \right>}

\DeclareMathOperator{\trace}{Tr}

\DeclareMathOperator{\Real}{Re}
\DeclareMathOperator{\Imaginary}{Im}

\DeclareMathAlphabet{\mathcalligra}{T1}{calligra}{m}{n}
\DeclareMathAlphabet{\mathpzc}{OT1}{pzc}{m}{it}

\begin{document}
\title{Quantum quenches of ion Coulomb crystals across structural instabilities II: \\
thermal effects.}
\author{Jens D.~Baltrusch$^{1,2}$}
\email[Email: ]{jens.baltrusch@physik.uni-saarland.de}
\author{Cecilia Cormick$^{1,3}$}
\author{Giovanna Morigi$^{1}$}
 \affiliation{
	$^1$Theoretische Physik, Universit\"at des Saarlandes, D-66123 Saarbr\"ucken, Germany\\
	$^2$Grup d'Optica, Departement de F\'isica, Universitat Aut\`onoma de Barcelona, E-08193 Bellaterra, Spain\\
	$^3$Institut f\"ur Theoretische Physik, Universit\"at Ulm, D-89069 Ulm, Germany
}
\date{\today
}
\begin{abstract}
We theoretically analyze the efficiency of a protocol for creating mesoscopic superpositions of ion chains, described in [Phys. Rev. A {\bf 84}, 063821 (2011)],  as a function of the temperature of the crystal. The protocol makes use of spin-dependent forces, so that a coherent superposition of the electronic states of one ion evolves into an entangled state between the chain's internal and external degrees of freedom. Ion Coulomb crystals are well isolated from the external environment, and should therefore experience a coherent, unitary evolution, which follows the quench and generates structural Schr\"odinger cat-like states.  The temperature of the chain, however, introduces a statistical uncertainty in the final state. We characterize the quantum state of the crystal by means of the visibility of Ramsey interferometry performed on one ion of the chain, and determine its decay as a function of the crystal's initial temperature. This analysis allows one to determine the conditions on the chain's initial state in order to efficiently perform the protocol. 
\end{abstract}
\pacs{03.65.Ud, 42.50.Dv}
%
\keywords{Ion Coulomb Crystals, Structural Superpositions States, Ramsey Interferometry}
\maketitle
\section{Introduction}
  \label{sec:introduction}
  
The quantum to classical transition is an intriguing problem of quantum physics~\cite{Zurek:2003} and a central issue of quantum-based technologies, where efforts are being invested in developing protocols for implementing quantum dynamics of systems of increasing size~\cite{QT:size}. Increasing the size of a system is usually associated with loss of coherence: Even when the physical object undergoes unitary evolution, the particles composing it can often be seen as a reservoir for each individual one~\cite{FordKacMazur}. As a result, a coherent and localized excitation can dephase on a rate which increases with the number of components. Such dephasing can be examined using the so-called Loschmidt echo~\cite{Zanardi,Paz,Cormick}, which can be measured by means of the visibility of an interferometric measurement performed on the system~\cite{Rossini,DeChiara:2008}.

\begin{figure}[btp]
  \centering
  \phantom{(a)}\includegraphics[width=0.48\textwidth]{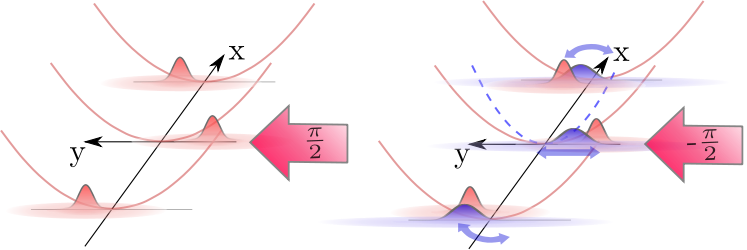}\put(-200,-10){(a)} \put(-65,-10){(b)} 
 \caption{(Color online) Ramsey interferometry with an ion chain whose vibrations are at temperature $T$. A quench across the linear-zigzag instability is performed by exciting the central ion with a laser pulse in presence of spin-dependent forces. In (a) the collective motion is initially in a thermal state of a zigzag structure, and the central ion in the internal state $|g\rangle$. A $\uppi/2$ laser pulse prepares it in the superposition $ (|g\rangle + |e\rangle)/\sqrt{2}$. (b) The ion in state $ |e\rangle$ experiences a tighter state-dependent potential. The corresponding conditional dynamics entangle the ions' internal and external degrees of freedom.  A subsequent laser pulse performs a $-\uppi/2$~rotation on the ion's internal transition. The final occupation of the ground state $|g\rangle$ as a function of the time $t$ elapsed between the two pulses allows one to extract information on quantum coherence and entanglement created by these dynamics. }
  \label{fig:1}
\end{figure}
In crystals of trapped ions such an interferometric measurement can be performed on an internal transition of one ion of the crystal. In Ref.~\cite{DeChiara:2008} a protocol for implementing Ramsey interferometry on the ion of the crystal has been proposed. It was shown that the visibility of the interferometric signal, here corresponding to the occupation of one electronic state of the ion, gives information on the quantum state of the crystal when analysed as a function of the time $t$ elapsed between the two Ramsey pulses.  This protocol is at the basis of the proposal of Ref.~\cite{Baltrusch:2011}    to create a superposition of two different crystalline structures across the linear-zigzag structural transition~\cite{Birkl:1992,Raizen:1992,Fishman:2008}: The mesoscopic superposition of a crystal in the linear and in the zigzag structure can be accessed by driving the electronic transition of one ion of the chain in a set-up where the trap frequency depends on the electronic state~\cite{Baltrusch:2011, Li-Lesanovsky}.  In these settings, a first laser pulse prepares the ion in a coherent superposition of the electronic states, which evolves into an entangled state between the chain's internal and external degrees of freedom  as sketched in Fig.~\ref{fig:1}. The visibility of the Ramsey signal does indeed decay fast with the time elapsed between the pulses, even when the chain is prepared at zero temperature and the dynamics are purely unitary. However,  for quenches close to the structural instability the visibility exhibits quasiperiodic revivals which are visible at longer times $t$. These revivals appear with a frequency which is determined by the frequency of the zigzag mode and persist when the size of the crystal is increased. The analysis of the spectrum of the visibility's temporal behaviour shows features that can be associated with the presence of entanglement generated by the first quantum quench~\cite{Baltrusch:2012}. 

The studies in Refs.~\cite{Baltrusch:2011,Baltrusch:2012} assume unitary dynamics of a chain initially at $T=0$. The assumption of unitary dynamics is reasonable for state-of-the-art ion-trap experiments \cite{Leibfried:2003,HaeffnerRoosBlattPhysRep2008,Schaetz:2012a}, in which the coherence times are of the order or larger than the typical time scales required for observing the dynamics predicted in Ref.~\cite{Baltrusch:2011}. Much more stringent is, however, the condition that the chain should be initially at temperature $T=0$. A feasibility analysis needs in fact a more quantitative statement on the temperatures required so that the protocol can be successfully performed.  The present work extends the analysis of Ref.~\cite{Baltrusch:2012} to the case in which the chain is initially at finite temperature~$T$, considering temperatures that can be achieved by means of Doppler or ground-state cooling \cite{Eschner:2003}. The behaviour of the visibility as a funtion of $T$ is analysed for small quenches across the linear-zigzag phase transition. This allows us to determine the experimental requirements on the temperature of the chain in order to perform the protocol. 

The article is organized as follows: In Sec.~\ref{sec:2} the proposal of Ref.~\cite{Baltrusch:2011} is summarized. The theoretical model is  presented in Sec.~\ref{sec:3}, which includes the detailed evaluation of the visibility signal as a function of the number of ions and of the initial temperature of the crystal. The behaviour of the visibility is analysed in  Sec.~\ref{sec:4} for a chain of three ions and different temperatures, and the conclusions are drawn in Sec.~\ref{sec:5}. Theoretical details for the derivation of the results in Sec.~\ref{sec:3} are given in the appendix.

\section{Ion Coulomb crystals in a thermal state}
  \label{sec:2}

In this section we briefly review the physical model and the protocol proposed in Ref.~\cite{Baltrusch:2011}. These are the starting points of the analysis performed in the following sections. 
\subsection{Ion Coulomb Crystals in State-dependent Traps}

The system we consider are atomic ions, which are confined in an external anisotropic potential. The ions have been laser cooled to sufficiently low temperatures, so that they perform small vibrations about the equilibrium positions determined by the competition between the external trap and the Coulomb repulsion~\cite{Dubin:1999,Morigi:2004}. 

In the following we assume $N$ identical ions with mass~$m$ and charge~$q$, which are confined by a linear Paul trap~\cite{Raizen:1992} or a Penning trap~\cite{Thompson:2002}. The ions are assumed to have been prepared at low temperature $T$ by means of laser cooling, such that they are aligned along a string and perform small vibrations about their equilibrium positions~\cite{Eschner:2003}. In this work we assume that after laser cooling all ions are in the same stable electronic state, which is denoted by $\ket{g}$. In absence of external perturbations, their motion is governed by the Hamiltonian~\cite{Morigi:2004,Baltrusch:2012}
\begin{equation}
 H_g =E_0^g+\sum_{j=1}^{3N}  \hbar \omega^{g}_j \left( b^{g}_j{}^\dagger b^{g}_j+\frac{1}{2}\right) \,.
\end{equation}
This Hamiltonian describes the harmonic vibrations about the equilibrium positions along the string, where $E_0^g$ is the energy of the classical ground state, $\omega^{g}_j$ are the normal mode frequencies of the crystal forming a linear (or a zigzag) chain and  $b^{g}_j{}^\dagger,  b^{g}_j$ are the corresponding bosonic creation and annihilation operators.

Structural superpositions can be obtained by preparing, for instance, one ion in a coherent superposition of state $\ket{g}$ and a second stable state $\ket{e}$, in which the ion experiences an additional potential due to a dipole trap~\cite{Schaetz:2010,Schaetz:2012b,Drewsen:2012,Baltrusch:2011}. When this potential is sufficiently steep, the ion's equilibrium position is displaced with respect to the case in which all ions are in state $\ket{g}$. Then, the long-range Coulomb repulsion causes a distortion of the crystalline structure, such that the ground state and the normal mode spectrum are now different. Therefore, when one ion is in $\ket{e}$, its dynamics are governed by the Hamiltonian
\begin{equation}
 H_e =E_0^e+\sum_{j=1}^{3N}  \hbar \omega^{e}_j \left( b^{e}_j{}^\dagger b^{e}_j+\frac{1}{2}\right) \,.
\end{equation}
Here, $E_0^e$ is the energy of the corresponding classical ground state, $\omega^{e}_j$ are the normal mode frequencies and  $b^{e}_j{}^\dagger,  b^{g}_j$ are the corresponding bosonic creation and annihilation operators. By preparing the ion's internal state  in a linear superposition of $\ket g$ and $\ket e$, the dynamics generated by the state-dependent Hamiltonian entangle the electronic and motional degrees of freedom~\cite{Baltrusch:2011}. A probe of quantum coherence and entanglement can be obtained by performing Ramsey interferometry with a single ion of the crystal~\cite{DeChiara:2008}. The scheme is sketched in the following. 

\subsection{Ramsey Interferometry with Thermal States}

In Ref.~\cite{DeChiara:2008} it was proposed to use Ramsey interferometry as a tool to probe the dynamics and thermodynamics of ion chains close to the zigzag instability. In Ref.~\cite{Baltrusch:2011} it was shown that the visibility of the Ramsey signal in chains of three ions presents features that can be associated with the creation of a superposition of motional states corresponding to linear and zigzag structures. In Ref.~\cite{Baltrusch:2012} it was shown that some of these features are found for chains of generic size~$N$. In these articles, it was assumed that the chain was prepared in the vibrational ground state, at $T=0$. The scope of this paper is to analyze how the coherence of the superposition is affected when the initial vibrational state is not pure but, say, a thermal state, as it is the experimental situation after laser cooling the chain.
In this subsection we briefly review the interferometric scheme and derive the expression for the visibility, which is going to be explicitly evaluated in the rest of this work.

Let the initial state of the ion chain be described by the density matrix $ \varrho_0 = \varrho(t=0)$. This reads 
\begin{equation}
\label{initial:state}
  \varrho_0 = \ketbra{g}{g} \otimes \rho_0 \,,
\end{equation}
where 
\begin{equation*} 
	\rho_0=\frac{1}{Z}\exp\left(-\frac{H_g}{k_BT}\right)\,, 
\end{equation*}
is the density matrix for the external degrees of freedom, with $k_B$ Boltzmann constant and $Z={\rm Tr}\{{\rm e}^{-H_g/(k_BT)} \}$ the partition function. In Fig.~\ref{fig:1}a the initial state is taken to be a thermally excited zigzag structure. A laser pulse applied for a time $\Delta\tau$ drives resonantly the transition $\ket{g}\to\ket{e}$ of the central ion, which we label by $j_0$. Assuming that the pulse area corresponds to a $\uppi/2$ rotation of the dipole, while its duration $\Delta\tau$ is sufficiently short, so that the chain motion can be neglected during  $\Delta\tau$, then the density matrix immediately after the pulse takes the form  
\begin{equation*}
\varrho_1=U\varrho_0 U^\dagger\,,
\end{equation*}
where $U=(\ketbraIX{e}{j_0}{e}+\ketbraIX{g}{j_0}{g}+R_{\vek{k}} \ketbraIX{e}{j_0}{g}+R_{\vek{k}}^\dagger \ketbraIX{g}{j_0}{e})/ \sqrt{2}$ is the evolution operator describing the dynamics due to the laser pulse and the operator $R_{\vek{k}} (\vek{x})= \mathrm{e}^{\mathrm{i} \vek{k}\cdot \vek{x}}$ describes the mechanical effect on the crystal associated with the absorption of a laser photon. 

The crystal evolves then freely for a time $t$ according to the Hamiltonian $H=\ketbra{e}{e}H_e+\ketbra{g}{g}H_g$, such that the density matrix at time $t$ reads $\varrho(t)=\varrho_2$, with
\begin{equation*}
\varrho_2=\sum_{p,q=e,g}\rho_{2,pq}\ketbra{p}{q}\,,
\end{equation*}
and
\begin{align}
\rho_{2,ee}(t) &= \frac{1}{2}  U_e(t) R_{\vek{k}} \rho_0 R_{\vek{k}}^\dagger U_e^\dagger(t) \,,\\
\rho_{2,gg}(t) &= \frac{1}{2} U_g(t) \rho_0 U_g^\dagger(t)   \,,\\
\rho_{2,eg}(t) &=\rho_{2,ge}^\dagger(t)  =
      \frac{\mathrm{e}^{\mathrm{i}\phi} }{2} U_e(t) R_{\vek{k}} \rho_0 U_g^\dagger(t) \,.
\end{align}
There, $U_p(t)=\exp(-\mathrm{i}H_pt/\hbar)$, while $\phi$ is a phase-shift applied when the atom is in the excited state, which allows one to perform interferometry. At this stage, atomic motion and internal degrees of freedom are entangled by the state-dependent evolution. A graphical representation is shown in Fig.~\ref{fig:1}b, illustrating a coherent superposition between zigzag and linear chain. Information about this structural superposition can be extracted by measuring the probability that the central ion is in the ground state after a second pulse has been applied which performs a $-\uppi/2$-pulse. The probability that the ground state of the ion is occupied after the pulse reads
\begin{equation}
\label{Prob}
 \mathcal P_g (\phi,t) = \trace \{ \varrho_f \ketbraIX{g}{j_0}{g} \} = \trace \{ \rho_{f,gg}(t) \} \,,
\end{equation}
where $\varrho_f$ is the density matrix after the second pulse and 
 \begin{multline}
 \label{rho:f}
  \rho_{f,gg}(t) = \frac{1}{2}  \Bigl(  
	    \rho_{2,gg}(t) +R_{\vek{k'}}^\dagger \rho_{2,ee}(t) R_{\vek{k'}}  \\
	  + \mathrm{e}^{\mathrm{i}\phi} \rho_{2,ge}(t) R_{\vek{k'}}
	  + \mathrm{e}^{-\mathrm{i}\phi} R_{\vek{k'}}^\dagger \rho_{2,eg}(t) 
	   \Bigr) \,,
  \end{multline}
with ${\bf k'}$ the wave vector of the second pulse. Using Eq.~\eqref{rho:f} in Eq.~\eqref{Prob}, the probability can be recast in the form
\begin{align}
\mathcal P_g (\phi,t) 
&= \frac{1}{2} \left( 1 + \Real \left[ \mathrm{e}^{\mathrm{i}\phi} \mathcal O(t) \right]   \right)
\,,
\end{align}
where $\mathcal O(t) $ measures the coherence between ground and excited state. It reads
\begin{equation}
  \label{eq:mixedoverlap}
  \mathcal O (t) =  \trace \left\{ R_{\vek{k'}}^\dagger U_e R_{\vek{k}} \rho_0 U_g{}^\dagger \right\}  \,,
\end{equation}
and determines the visibility $\mathcal V$ of the Ramsey signal through the relation $\mathcal V=|\mathcal O|$. In Sec.~\ref{sec:3} we determine $\mathcal O(t) $ as a function of the time elapsed between the two pulses and of the initial temperature of the chain, and in Sec.~\ref{sec:4} we discuss its behavior as a function of the temperature $T$ for a chain of three ions in a two-dimensional geometry.

\section{Evaluation of the visibility of the Ramsey fringes}
\label{sec:3}
In this section we carry out the theoretical evaluation of the visibility of the Ramsey signal as a function of the temperature~$T$ and of the time~$t$ elapsed between the two Ramsey pulses. The calculation here reported extends the one presented in~\cite{Baltrusch:2012}, which was performed assuming that the ion crystal is initially prepared in the vibrational ground state. We also include the possibility of a mechanical effect associated with the absorption and emission of a photon of the the pulses. The final result is reported in Eq.~\eqref{eq:visibility:final}. It is valid for a three-dimensional geometry, for any number of ions and for any initial temperature~$T$, as long as the assumption of that the ions perform harmonic vibrations about their equilibrium positions is valid.

\subsection{Some useful relations}

In order to evaluate Eq.~\eqref{eq:mixedoverlap}, we make use of the unitary transformations which relate the states and normal-mode operators of the two structures. These have been derived in Ref.~\cite{Baltrusch:2012}, and are reported in this section. 

We write the ions' positions $r_j$ ($j=1,\ldots,3N$) as small excursions $q^g_j, q^e_j$ away from the equilibrium positions of the corresponding structures, $r_j^{g}, r_j^{e}$, namely,
\begin{equation}
\label{eq:ansatzlink}
r_j = r_j^{g} + q_j^{g} =  r_j^{e} + q_j^{e} \,,
\end{equation}
where the superscripts $g$ and $e$ indicate whether the central ion is in the ground state~$\ket{g}$ or in the excited state~$\ket{e}$. We denote by $d_j^g=r_j^e-r_j^g$ the equilibrium displacements of ion $j$ between the structures. The normal modes are obtained by diagonalizing Hamiltonian~$H_s$ ($s=g,e$), after it has been expanded around the equilibrium positions up to second order in the displacements~$q_j^s$. Thus, they depend  on the internal state of the central ion and read
\begin{equation}
\label{Eq:Q}
 Q_l^{ s} = \sum_k \mathbf M_{kl}^{s} q_k^{s}\,,
\end{equation} 
where matrix $\mathbf M^s$ is the orthogonal transformation diagonalizing the harmonic part of the potential; details are reported in Ref.~\cite{Baltrusch:2012}. Equations~\eqref{eq:ansatzlink} and~\eqref{Eq:Q} give the mapping connecting the normal modes in the two structures:
\label{eq:linkmodes}
\begin{align}
\label{eq:linkmodes:Q}
Q_j^{g} &= \sum_k \mathbf{T}_{jk} Q_k^{e} + D^g_j  \,,\\
\label{eq:linkmodes:P}
P_j^{g} &= \sum_k \mathbf{T}_{jk} P_k^{e} \,,
\end{align}
with $P_j^s$ the momentum canonically conjugated to the displacement $ Q_l^{ s}$ and 
\begin{align}
  \mathbf{T}_{jl}&=\sum_k \mathbf{M}_{kj}^{g} \mathbf{M}_{kl}^{e} \,, &
  D_j^g &= \sum_k \mathbf{M}_{kj}^{g} d_k^g\,.
\end{align}
Correspondingly, the annihilation and creation operators ${b_j^s}$ and  ${b_j^s}^\dagger$, defined by relations $b_j^s= \sqrt{m\omega^s_j/(2\hbar)} [Q_s + {\rm i}P_s/(m\omega^s_j)]$ and its adjoint, are related by the Bogoliubov transformations
\begin{subequations}
\begin{align}	
\label{eq:bogoliubov}
b_j^{g}{}^{\phantom\dagger} &= \sum_k u_{jk} b^{e}_k{}^{\phantom\dagger} - \sum_k v_{jk} b^{e}_k{}^\dagger + \beta^g_j \,,\\
\label{eq:inverseBogoliubov}
    b_j^{e}{}^{\phantom\dagger} &= \sum_k u_{kj} b^{g}_k{}^{\phantom\dagger} + \sum_k v_{kj} b^{g}_k{}^\dagger + \beta^e_j \,,
\end{align}
\end{subequations}
with the real and dimensionless coefficients 
\begin{subequations}
\label{eq:bogoliubov:coeffs}
 \begin{align}
\label{eq:bogoliubov:u}
u_{jk} &= \frac{\mathbf{T}_{jk}}{2}  \left[ \sqrt{\frac{\omega_k^{e}}{\omega_j^{g}}} + \sqrt{\frac{\omega_j^{g}}{\omega_k^{e}}} \right]  \,,\\
\label{eq:bogoliubov:v}
v_{jk} &= \frac{\mathbf{T}_{jk}}{2}  \left[ \sqrt{\frac{\omega_k^{e}}{\omega_j^{g}}} - \sqrt{\frac{\omega_j^{g}}{\omega_k^{e}}}\right] \,,
\end{align}
\end{subequations}
and displacements $\beta^g_j = \sqrt{m \omega_j^{g}/2\hbar} \, D^g_j$, such that
\begin{subequations}
  \label{eq:mapping:phasespacedisplacements}
  \begin{align}
    \beta^e_j &= - \sum_k \sqrt{\frac{\omega^e_j}{\omega^g_k}} \mathbf{T}_{kj} \beta^g_k = - \sum_k (u_{kj} + v_{kj}) \beta^g_k \,, \\
    \beta^g_j &= - \sum_k \sqrt{\frac{\omega^g_j}{\omega^e_k}} \mathbf{T}_{jk} \beta^e_k = - \sum_k (u_{jk} - v_{jk}) \beta^e_k \,.
  \end{align}
\end{subequations}
The vibrational ground state of each structure is denoted by $\ket{0}_s$, with $s=g,e$. They are mapped into one another by the transformation~\cite{Baltrusch:2012,FetterAoP72}
\begin{equation}
  \ket 0_{g} = Z \, \mathcal D_e(\beta^e_1,\dotsc,\beta^e_{3N}) \, e^\mathrm A \ket 0_{e} \,.
\end{equation}
Here, $\mathrm A$ reads
\begin{equation}
\mathrm A = \frac{1}{2}\sum_{jk} A_{jk} b^e_j{}^\dagger b^e_k{}^\dagger
\end{equation}
with $  A_{jk}$ a real and symmetric matrix with elements 
  \begin{equation}
  A_{jk} = \sum_l (u^{-1})_{jl} v_{lk} \,.
 \end{equation}
 The scalar
\begin{equation}
   \label{eq:mapping:Z}
  Z = \det \left[ \left( 1-\mathrm A^2 \right)^{1/4} \right] \,
\end{equation}
is warranting the correct normalization, while 
the term $\mathcal D_e(\beta^e_1,\dotsc,\beta^e_{3N})$ is a displacement operator of the $3N$ normal modes, when the structure is the one corresponding to the central ion being excited. It is defined as
\begin{equation}
{\mathcal D}_e(\beta^e)=\otimes_{j=1}^{3N}{\mathcal D}_e^{(j)}(\beta^e_j)\,,
\end{equation}
with 
\begin{equation}
{\mathcal D}_s^{(j)}(\beta^e_j)=\exp \{\beta^e_j b_s^\dagger - \beta^e_j{}^* b_s) \}\,.
\end{equation}
It is useful to introduce the relation between displacement operators in the basis of normal modes of each structure. For a generic displacement $\lambda^g$, here given for the structure in which the central ion is in the ground state, they are related by the equation
\begin{equation}\label{map:displacements}
  \mathcal D_g(\lambda^g) = \mathrm{e}^{\mathrm{i}\varphi[\lambda^g]} \mathcal D_e(\lambda^e) \,,
\end{equation}
where
\begin{equation}\label{eq:phase:lambda}
  \varphi[\lambda^g]= 2\Imaginary \Bigl[ \sum_j  \lambda^g_j\beta^g_j \Bigr] \,,
\end{equation}
and
\begin{equation}
\lambda^e_j = \sum_l (\lambda^g_l u_{lj} + \lambda^g_l{}^* v_{lj}) \,.
\end{equation}

\subsection{Evaluation of the visibility for any initial state}
\label{sec:overlap:evaluation}
We now evaluate Eq.~\eqref{eq:mixedoverlap}, whose modulus is the visibility for an arbitrary initial state. We use that $R_{\vek{k}}({\vek{x}})=\exp({\mathrm{i}}{\vek{k}}\cdot {\vek{x}})$ is a displacement operator for each normal mode, such that $R_{\vek{k}}({\vek{x}})={\mathcal D}_e(\kappa)$. 
Here,
\begin{equation}
\kappa_j = \mathrm{i} \sqrt{\frac{\hbar}{2m\omega^e_j}} K_j \,,
\end{equation}
where $K_j = \left( k_x \mathbf{M}^e_{j_{0x},j} + k_y \mathbf{M}^e_{j_{0y},j} + k_z \mathbf{M}^e_{j_{0z},j} \right)$ is the projection of the wave vector onto the normal mode~$j$, assuming that ion $j_0$ is illuminated  ($j_{0\alpha}$ labels the $\alpha=x,y,z$ displacement of the ion).

It is convenient to use a coherent state basis for performing the evaluation  of Eq.~\eqref{eq:mixedoverlap}. Therefore, we take the trace in Eq.~\eqref{eq:mixedoverlap} over the basis of coherent states $\ket{\alpha}_e$ of the harmonic oscillators corresponding to the normal modes when the ion is in state $e$, where $\ket{\alpha}_e=\otimes_j \ket{\alpha_j}_e$, such that ${\mathcal D}_e^{(j)}(\alpha_j)\ket{0}_e=\ket{\alpha_j}_e$. Using the cyclic properties of the trace, we recast Eq.~\eqref{eq:mixedoverlap} in the form
\begin{align}\label{eq:overlap:evaluation:trace}
  \mathcal O(t) = \int \frac{\mathrm d^{6N}\alpha}{\uppi^{3N}} \matrixElementSub{\alpha}{e}{   R_{\vek{k}} \rho_0 U_g{\!\!}^\dagger{\,} R_{\vek{k'}}{\!\!\!\!}^\dagger{\,\,} U_e}{\alpha}{e} \,.
\end{align}
The initial density matrix can be expressed in the form 
\begin{equation} \label{densitymatrixP}
 \rho_0 =   \int \frac{\mathrm d^{6N}\lambda^g}{\uppi^{3N}} \, P_0(\lambda^g)  \ketbraIX{\lambda^g}{g}{\lambda^g}\,,
\end{equation}
where  $\ket{\lambda^g}_g=\otimes_j \ket{\lambda^g_j}_g$ is the basis of coherent states of the harmonic oscillators, corresponding to the normal modes when the ion is in state $g$, and $P_0(\lambda^g)$ is the Glauber-Sudarshan-$P$ distribution containing the information over the initial state~\cite{Carmichael1}, with $\lambda^g=(\lambda^g_1,\ldots,\lambda^g_{3N})$. Using Eq.~\eqref{densitymatrixP} in Eq.~\eqref{eq:overlap:evaluation:trace}, we find
\begin{eqnarray}
\label{Eq:O:coh}
  \mathcal O(t) &=& \int \frac{\mathrm d^{6N}\alpha}{\uppi^{3N}} \int \frac{\mathrm d^{6N} \lambda^g}{\uppi^{3N}} \, P_0(\lambda^g)  \\
  &\times&\matrixElementSub{ \alpha}{e}{R_{\vek{k}}}{ \lambda^g}{g} \matrixElementSub{\lambda^g}{g}{U_g{\!\!}^\dagger{\,} R_{\vek{k'}}{\!\!\!\!}^\dagger{\,\,} U_e }{\alpha}{e} \,.\nonumber
\end{eqnarray}
This expression contains two matrix elements. We write the first one as
\begin{equation}
\label{matrixEl1}
  \matrixElementSub{ \alpha}{e}{R_{\vek{k}}}{ \lambda^g}{g} = Z \mathrm{e}^{\mathrm{i}\varphi[\lambda^g]} \matrixElementSub{ \alpha}{e}{\mathcal D_e( \kappa) \mathcal D_e ( \lambda^e) \mathcal D_e ( \beta^e) {\mathrm e}^{\mathrm A}}{0}{e} \,,
\end{equation}
where we used $\ket{\lambda^g}_g = \mathcal D_g(\lambda^g)\ket{0}_g$. The second matrix element in the right-hand side of Eq.~\eqref{Eq:O:coh} can be rewritten as 
\begin{align}
 &\matrixElementSub{\lambda^g}{g}{U_g{\!\!}^\dagger{\,} R_{\vek{k'}}{\!\!\!\!}^\dagger{\,\,} U_e }{\alpha}{e} = \matrixElementSub{\lambda^g(t)}{g}{ R_{\vek{k'}}{\!\!\!\!}^\dagger{\,\,} }{\alpha(t)}{e} 
\nonumber \\
 =&Z \mathrm{e}^{-\mathrm{i}\varphi[\lambda^g(t)]} \matrixElementSub{0}{e}{ \mathrm{e}^{\mathrm A^\dagger} \mathcal D_e{\!\!}^\dagger ( \beta^e) \mathcal D_e{\!\!}^\dagger ( \lambda^e(t))  \mathcal D_e{\!\!}^\dagger(\kappa')  }{\alpha(t)}{e} \,,
 \label{matrixEl2}
\end{align}
where
\begin{equation}\label{eq:lambdat}
  \lambda^e_j(t) =  \sum_k \left( \lambda^g_k \mathrm{e}^{-\mathrm{i}\omega^g_k t} u_{kj} + \lambda^g_k{}^* \mathrm{e}^{+\mathrm{i}\omega^g_k t} v_{kj} \right) \,.
\end{equation}
Using these results, Eq.~\eqref{eq:overlap:evaluation:trace}  can be cast in the form  
\begin{multline}
\label{Eq:O:middle}
  \mathcal O(t) =  \int \frac{\mathrm d^{6N} \alpha}{\uppi^{3N}} \int \frac{\mathrm d^{6N} \lambda^g}{\uppi^{3N}}  \, Z^2 \mathrm{e}^{\mathrm{i}\varphi} P_0(\lambda^g) \\ 
\times  \matrixElementSub{ \alpha}{e}{\mathrm{e}^{\mathrm{A}(\theta)}}{ \theta}{e} \matrixElementSub{ \theta'}{e}{ \mathrm{e}^{\mathrm{A}^\dagger(\theta')}    }{\alpha(t)}{e} \,,
\end{multline}
where
\begin{equation}
  \varphi = \varphi[\lambda^g]-\varphi[\lambda^g(t)]  + \varphi_{\theta} - \varphi_{\theta'} \,,
\end{equation}
with 
\begin{subequations}
\begin{align}
  \varphi_{\theta} &= \Imaginary \Bigl[ \sum_j (\kappa_j + \beta^g_j) \lambda^e_j{}^* + \sum_j \kappa_j \beta^e_j\Bigr] \,, \\
  \varphi_{\theta'} &= \Imaginary \Bigl[ \sum_j (\kappa'_j + \beta^g_j)\lambda^e_j{}^*(t) + \sum_j \kappa'_j \beta^e_j\Bigr]\,,
\end{align}
\end{subequations}
and where $\kappa'$ is the displacement due to the emission of a photon with wave vector ${\bf k'}$. In Eq.~\eqref{Eq:O:middle} we have introduced the quantities 
\begin{subequations}\label{eq:combined_displacements}
  \begin{align}
    \theta_j &= \kappa_j + \beta^e_j + \lambda^e_j \,, \\
    \theta'_j &= \kappa'_j + \beta^e_j + \lambda^e_j(t) \,,
  \end{align}
\end{subequations}
as well as the operators
\begin{align}
 \mathrm A(\theta) &= \frac{1}{2} \sum_{jk} A_{jk} (b^e_j{}^\dagger - \theta_j{}^*)(b^e_k{}^\dagger - \theta_k{}^*) \,.
\end{align}
Exchanging the order of the integrations and evaluating the operators, Eq.~\eqref{Eq:O:middle} becomes
 \begin{equation}
 \mathcal O(t) = \int \frac{\mathrm d^{6N} \lambda^g}{\uppi^{3N}}  \, Z^2 \mathrm{e}^{\mathrm{i}\varphi} P_0(\lambda^g) \; \mathfrak I_{\alpha}(\lambda^g) \,,
\end{equation}
with 
\begin{align}
  \mathfrak I_{\alpha}(\lambda^g) =& \int \frac{\mathrm d^{6N} \alpha}{\uppi^{3N}}  
  \matrixElementSub{ \alpha}{e}{\mathrm{e}^{\mathrm{A}(\theta)}}{ \theta}{e}   \matrixElementSub{ \theta'}{e}{ \mathrm{e}^{\mathrm{A}^\dagger(\theta')}    }{\alpha(t)}{e} \nonumber\\
  =&  \int \frac{\mathrm d^{6N} \alpha}{\uppi^{3N}} 
    \braketIX{e}{\alpha}{\theta}{e}
    \braketIX{e}{\theta'}{\alpha(t)}{e} \matrixElementSub{ \alpha}{e}{\mathrm{e}^{\mathrm{A}(\theta)}}{ \alpha}{e}\nonumber\\
    &\hphantom{\int \frac{\mathrm d^{6N} \alpha}{\uppi^{3N}} } \times     \matrixElementSub{ \alpha(t)}{e}{\mathrm{e}^{\mathrm{A}^\dagger(\theta')}}{ \alpha(t)}{e} 
\label{eq:integral_in_alpha}
\end{align}
The explicit evaluation of the integral in the variables $\alpha$ is reported in Appendix~\ref{app:overlapintegral}, and leads to the expression 
\begin{equation}\label{eq:overlapintegral}
 \mathcal O(t) = \int \frac{\mathrm d^{6N} \lambda^g}{\uppi^{3N}} \, P_0(\lambda^g)  
    \frac{Z^2 \mathrm{e}^{\mathrm{i}\varphi}}{\sqrt{\det \Omega}}  \mathrm{e}^{G^*(\theta')}\mathrm{e}^{G(\theta)} \mathrm{e}^{\frac{1}{4} \vek{s}^T \Omega^{-1} \vek{s}} \,,
\end{equation}
where
 \begin{equation}
  \label{def:constant:G}
  G ( \gamma)=  \sum_{jk}  \frac{A_{jk}}{2} \gamma_j^*\gamma_k^*  - \sum_{j} \frac{|\gamma_j|^2}{2} \,,
\end{equation}
with $\gamma_j = \theta_j, \theta'_j$. Here, $\Omega$ is a complex symmetric $6N$-by-$6N$ matrix, which reads
\begin{equation}
 \Omega = 
	\begin{pmatrix}
	    \Omega^{++} & \Omega^{+-} \\ \Omega^{-+} & \Omega^{--}
        \end{pmatrix} =
	\begin{pmatrix}
	      1-\mathrm{A}^+ & \;\, -\mathrm{i}\mathrm{A}^- \\ -\mathrm{i}\mathrm{A}^- & 1+\mathrm{A}^+
	\end{pmatrix} 
\end{equation}
with 
\begin{equation}
    \mathrm{A}_{jk}^\pm =  \frac{1}{2} \bigl( A_{jk}(\mathrm{e}^{-\mathrm{i}(\omega_j^e+\omega_k^e)t} \pm 1 ) \bigr)   \,. 
\end{equation}
Moreover,  $\vek{s}$ is a $6N$-dimensional vector given by
\begin{equation}
  \vek{s} = 
	\begin{pmatrix}
		\phantom{-i} S^+ \\ -\mathrm{i}S^-
	\end{pmatrix} ,
\end{equation}
with
\begin{subequations}
\begin{align}\label{def:shift}
    S^\pm_j [\theta,\theta']&= S_{j}[\theta] \pm S_{j}^*[\theta'] \mathrm{e}^{-\mathrm{i}\omega_j^e t} \,,\\
  S_j[\gamma] &= \sum_{k} A_{jk} \gamma_k^* - \gamma_j \,.
\end{align}
\end{subequations}
Equation~\eqref{eq:overlapintegral} gives the visibility as a function of an arbitrary initial state, for an arbitrary number of ions $N$ and accounting for the mechanical effect associated with the absorption and emission of a photon of the laser pulse. 

\subsection{Visibility for an initial thermal state}
\label{sec:overlap:evaluation}

We now evaluate the visibility when the chain is initially in a thermal state, as in Eq.~\eqref{initial:state}; we need to integrate in Eq.~\eqref{eq:overlapintegral} over the variables $\lambda^g$ taking the distribution $P_0(\lambda^g)=\prod_jP_0(\lambda^g_j)$, such that~\cite{Carmichael1}
\begin{equation}
P_0(\lambda^g_j) = \frac{1}{\uppi \mean{n^g_j}} \exp \left[ -\frac{\modulus{\lambda^g_j}^2}{\mean{n^g_j}}\right] \,,
\end{equation}
with 
\begin{equation}
\label{eq:modeoccupation}
  \mean{n^g_j} =\mean{b^g_j{}^\dagger b^g_j} = \frac{\mathrm{e}^{-\hbar\omega^g_j/k_B T}}{ 1-\mathrm{e}^{-\hbar\omega^g_j/k_B T}}\,,
\end{equation}
the mean vibrational number of mode $b_j^g$. The integral in the variable $\lambda^g$ is a Gaussian integral and the resulting visibility reads:
\begin{equation}
\label{eq:visibility:final}
 \mathcal O(t) = \frac{Z^2 \mathrm{e}^{\mathrm{i}\tilde \varphi} e^{\mathcal C}}{\mean{n_1} \dotsm \mean{n_{3N}}} \frac{\exp\left\{ \frac{1}{4} {\mathcal{L}}{}^T \mathcal X^{-1} {\mathcal{L}}\right\}}{\sqrt{\det \Omega \det \mathcal{X}}}  \,.
\end{equation}
This expression is valid for any initial temperature $T$ and any number of ions, as long as the harmonic approximation at the basis of our model is valid. In Eq.~\eqref{eq:visibility:final} we have introduced a series of quantities in order to provide a compact form. These quantities are given here in order to make the presentation self-consistent. 

The prefactors contain two exponentials, whose exponents take the form
\begin{equation*}
\tilde \varphi=(\varphi[\kappa]-\varphi[\kappa^{\prime}])/2\,,
\end{equation*}
and 
\begin{equation*}
\mathcal C=G(\zeta)+G^*(\zeta')+\sum_{j,k=1}^{3N}\sum_{\alpha,\beta=\pm}\frac{S_j^\alpha[\kappa,\kappa^\prime][\Omega^{-1}]_{jk}^{\alpha\beta}S_k^\beta[\kappa,\kappa^\prime]}{4}\,,
\end{equation*}
where
\begin{align}
 \zeta_j &= \kappa_j + \beta^e_j \,, & \zeta'_j &= \kappa'_j + \beta^e_j \,.
\end{align}
The vector $\mathcal{L}$ is conveniently decomposed into three parts,
\begin{equation}
\mathcal{L} = \mathcal{I} + \mathcal{J} + \mathcal{K} \,.
\end{equation}
The first term on the right-hand side is given by
\begin{equation}
  \begin{pmatrix}
\mathcal{I}^1_j \\
\mathcal{I}^2_j 
  \end{pmatrix} =
  \begin{pmatrix}
{I}^1_j(\zeta^*) + {I}^2_j(\zeta') \mathrm{e}^{-\mathrm{i} \omega^g_j t}\\ 
{I}^2_j(\zeta^*) + {I}^1_j(\zeta') \mathrm{e}^{+\mathrm{i} \omega^g_j t}
  \end{pmatrix}
\end{equation}
where
\begin{subequations}
\begin{align}
  I^1_l(\zeta^*) &= \sum_{jk} v_{lj} A_{jk} \zeta_k^* - \frac{1}{2} \sum_{j} \left( v_{lj} \zeta_j + u_{lj} \zeta_j^* \right) \,, \\
  I^2_l(\zeta^*) &= \sum_{jk} u_{lj} A_{jk} \zeta_k^* - \frac{1}{2} \sum_{j} \left( u_{lj} \zeta_j + v_{lj} \zeta_j^* \right) \,.
\end{align}
\end{subequations}
The second term can be written as
\begin{equation}
  \begin{pmatrix}
 \mathcal J^1_k \\
 \mathcal J^2_k
  \end{pmatrix}
= 
  \begin{pmatrix}
	\beta^g_j (1-\mathrm{e}^{-\mathrm{i}\omega^g_j t}) + \frac{1}{2} \bigl( J^{+}_k(\kappa) -J^{+}_k(\kappa') \mathrm{e}^{-\mathrm{i} \omega^g_k t}\bigr) \\
	\beta^g_j (\mathrm{e}^{+\mathrm{i}\omega^g_j t}-1) + \frac{1}{2} \bigl( J^{-}_k(\kappa) -J^{-}_k(\kappa') \mathrm{e}^{+\mathrm{i} \omega^g_k t}\bigr) 
  \end{pmatrix}
\end{equation}
with
\begin{equation}
 J^{\pm}_k(\kappa)  = \sum_j \Bigl( \kappa_j (u_{kj} + v_{kj}) \pm \beta^e_j (u_{kj} - v_{kj}) \Bigr)  
\end{equation}
The third term reads
\begin{align}
  \begin{pmatrix}
 \mathcal K^1_k \\
 \mathcal K^2_k
  \end{pmatrix} 
= \sum_{\alpha\beta} \sum_{jk} &
  \left[ 
\begin{pmatrix}
     \mathrm{Y}_{jl}  [\Omega^{-1}]^{\alpha\beta}_{jk} S^\beta_k[\kappa,\kappa']\\ 
     \mathrm{Y}^{\alpha}_{jl}  [\Omega^{-1}]^{\alpha\beta}_{jk} S^\beta_k[\kappa,\kappa'] 
\end{pmatrix} 
+ \right. \nonumber \\
& + 
 \left.
  \begin{pmatrix}
      S^\alpha_j[\kappa,\kappa'] [\Omega^{-1}]^{\alpha\beta}_{jk} \mathrm{Y}_{kl}  \\
      S^\alpha_j[\kappa,\kappa'] [\Omega^{-1}]^{\alpha\beta}_{jk} \mathrm{Y}^{\beta}_{kl} 
  \end{pmatrix} 
 \right] \,,
\end{align}
where
\begin{align}
  \mathrm{Y}_{jl} &= \sum_k A_{jk} v_{lk} -  u_{lj} \,,\\
  \mathrm{Y}_{jl}^{\pm}	  &= \pm \mathrm{Y}_{jl} \mathrm{e}^{-\mathrm{i}(\omega^e_j - \omega^g_l)t} \,.
\end{align}

The matrix $\mathcal X$ in Eq.~\eqref{eq:visibility:final} is given by the following expression,
\begin{align}
&
  \begin{pmatrix}
    \mathcal X_{lm}^{11} & 
    \mathcal X_{lm}^{12} \\ 
    \mathcal X_{lm}^{21} & 
    \mathcal X_{lm}^{22}  
  \end{pmatrix}
= 
  \begin{pmatrix}
   0 & 
   0 \\ 
   \mathcal T_{lm} &
   0  
  \end{pmatrix}
+
  \begin{pmatrix}
    \mathrm{Y}^0_{lm} & 
    -\frac{1}{2} \mathrm{e}^{-\mathrm{i}(\omega^g_l-\omega^g_m) t} \\
    -\frac{1}{2} & 
    \mathrm{Y}^0_{lm} \mathrm{e}^{+i(\omega^g_l+\omega^g_m) t}
  \end{pmatrix} \nonumber\\
&+
 \sum_{\alpha\beta} \sum_{jk} 
      \begin{pmatrix}
		\mathrm{Y}_{jl} & 0 \\ 0 &  \mathrm{Y}^\alpha_{jl}
      \end{pmatrix}\!
      \begin{pmatrix}
		[\Omega^{-1}]^{\alpha\beta}_{jk} &
		[\Omega^{-1}]^{\alpha\beta}_{jk} \\ 
		[\Omega^{-1}]^{\alpha\beta}_{jk} & 
		[\Omega^{-1}]^{\alpha\beta}_{jk}
      \end{pmatrix}\!
      \begin{pmatrix}
		\mathrm{Y}_{km} & 0 \\ 0 &  \mathrm{Y}^\beta_{km}
      \end{pmatrix}\! ,
\end{align}
with
\begin{equation}
\mathrm{Y}^0_{lm} = \frac{1}{2}\sum_j v_{lj} \mathrm{Y}_{jm} \,,
\end{equation}
and the thermal excitation, 
\begin{equation}
\mathcal T_{lm} = \delta_{lm} \mean{n^g_l}^{-1} \,.
\end{equation}

The integration in $\lambda^g$ is facilitated by changing to real and imaginary parts of $\lambda^g_j = x_j + \mathrm{i} y_j$, thereby introducing
\begin{equation}
  \begin{pmatrix}
    \mathcal{X}_{lm}^{xx} &
    \mathcal{X}_{lm}^{xy} \\
    \mathcal{X}_{lm}^{yx} &
    \mathcal{X}_{lm}^{yy}
  \end{pmatrix} =
  \begin{pmatrix}
    1 & \phantom{-}1 \\ \mathrm{i} & -\mathrm{i}
  \end{pmatrix}
  \begin{pmatrix}
    \mathcal X_{lm}^{11} & 
    \mathcal X_{lm}^{12} \\ 
    \mathcal X_{lm}^{21} & 
    \mathcal Y_{lm}^{22}  
  \end{pmatrix}
  \begin{pmatrix}
    1 & \phantom{-}\mathrm{i} \\ 1 & -\mathrm{i}
  \end{pmatrix}
\end{equation}
and
\begin{equation}
  \begin{pmatrix}
\mathcal{L}^x_j \\
\mathcal{L}^y_j
  \end{pmatrix} = 
  \begin{pmatrix}
\mathcal{L}^1_j + \mathcal{L}^2_j \\
\mathcal{L}^1_j - \mathcal{L}^2_j 
  \end{pmatrix}  \,. 
\end{equation}

\section{Results}
\label{sec:4}
We now analyze the visibility of the Ramsey fringes when the central ion is subject to a sequence of two Ramsey pulses in presence of a state-dependent potential. The results we present are obtained by evaluating explicitly the visibility in Eq.~\eqref{eq:visibility:final} for a given set of parameters, assuming that the vibrations along the direction perpendicular to the plane of the zigzag are frozen out, namely, the motion is effectively confined to the $x-y$ plane. We will focus on a chain composed by three ions in a linear trap with axial frequency $\nu_x$ and transverse secular frequency $\nu_y$. In the following we will consider values of $\nu_y$ close to the critical value $\nu_c$, separating the linear from the zigzag phase~\cite{Morigi:2004,Fishman:2008}, and use the dimensionless parameter 
\begin{equation}
g=\frac{\nu_y^2-\nu_{c}^2}{\nu_{c}^2}
\end{equation}
in order to indicate whether the ions form a linear array ($g>0$), or a zigzag chain ($g<0$). The instability is at $g=0$. The effect of the quench on the chain, due to the internal excitation of the central ion, is represented by a shift of the trapping frequency that the central ion experiences, denoted by $\nu_{\mathrm{dip}}$. The strength of the quench is here described by the dimensionless parameter 
\begin{equation}
\Delta =\frac{\nu_{\mathrm{dip}}^2}{\nu_{c}^2}\,,
\end{equation}
that is here taken to be positive, $\Delta>0$. Hence, when the central ion is in the excited state, the trapping potential it experiences is steeper. Table~\ref{Table:1} reports the experimental parameters corresponding to the values of $g$ and $\Delta$ we consider in this section.

\begin{table}[b!t]
 \centering
\begin{tabular*}{.45\textwidth}{c|c|c|c|c|c}
$g $	& -0.1 & -0.005 & 0  & 0.02 &\\
\hline
$\nu_y/(2\uppi)$ (\si{\mega\hertz}) & 1.470  & 1.545  & 1.549 & 1.565 & \\
\hline 
\hline
$\Delta $	& 0.005 & 0.01 & 0.015  & 0.02 & 0.025 \\
\hline
$\nu_{\mathrm{dip}}/(2\uppi)$ (\si{\kilo\hertz} &   110 & 155  & 190 & 219 & 245
\end{tabular*} 
\caption{\label{Table:1}Conversion table for the dimensionless quantities to actual frequencies used for three ions with an axial trap frequency of $\nu_x = 2\uppi \times \SI{1}{\mega\hertz}$. The critical frequency is $\nu_c=\sqrt{12/5}\nu_x=2 \uppi \times \SI{1.549}{\mega\hertz}$. }
\end{table}
The plots we present display the visibility, namely, the absolute value of the overlap $\mathcal O(t)$ in Eq.~\eqref{eq:visibility:final}, as a function of the time $t$ elapsed between the two pulses and of the temperature. The plots are evaluated for a chain composed by three $^9$Be$^+$ ions, at different values of $g$ and $\Delta$ and at different initial temperatures $T$ of the chain.  In the first part of this section we discard possible mechanical effects of the laser pulse; this situation can be realised with suitably tailored excitation schemes, for instance by taking copropragating  laser beams in a Raman scheme~\cite{Leibfried:2003}, or by using radiofrequency fields~\cite{Balzer:2006}. In the last part we then consider a pulsed excitation in which the mechanical effect is relevant and analyse its effect over the visibility signal. 

Before we start, some considerations on the choice of the parameter $\Delta$ are in order. We first note that the model we consider, a crystalline structure where the ions perform harmonic vibrations about the equilibrium conditions, require that anharmonicities are not relevant for the dynamics we investigate. This sets in general  an upper bound to the choice of the quench's amplitude $\Delta$. In addition, anharmonic corrections are naturally relevant very close to the linear-zigzag instability~\cite{Fishman:2008}, so that the initial and final state should be sufficiently distant from the critical point. Hence, this sets a lower bound to $\Delta$ when the quench is performed across the linear-zigzag instability, such that the initial state is, say, a zigzag structure and the excited state is a linear array. The parameters we choose are chosen in accordance with these conditions.

\subsection{Initial thermal excitation}
\label{sec:results:thermalexcitation}
We assume that the initial state of the crystal is a thermal state of the corresponding equilibrium structure at a given temperature $T$. Table~\ref{Table:2} reports the mean vibrational number of each normal mode for the values of $g$ and $T$ we consider in this section. 
In the following we will see that one normal mode will become important in our discussion. For the crystal being in the linear structure, this mode is (for the parameters considered) the zigzag mode~\cite{DeChiara:2008,Fishman:2008}. Its frequency and its motional pattern are displayed in Table~\ref{Table:2} in the top row in the upper block. The frequency of the zigzag mode goes to zero when approaching the linear-zigzag transition, and when the mode crosses the transition it becomes mixed with a second normal mode. The motional pattern is displayed for two values in the zigzag in the top row of the lower two blocks of Table~\ref{Table:2}. 
For convenience we will name in the following also this mode as the zigzag mode when the crystal is below the transition. We also will also use the term soft mode for this mode.

\begin{table}[bt]
 \centering 
\begin{tabular}{|r|c|r|r|r|r|}
\hline
  $\omega^g_j/2\uppi $ & mode & \multicolumn{4}{c|}{ $T\, (\mu\si{\kelvin})$ }\\
 $(\si{\mega\hertz})$ & & 5 & 10 & 50 & 100 \\
\hline
\hline
\multicolumn{6}{|c|}{$g=0.02$}\\
\hline
  0.2191 & $[\downarrow \uparrow \downarrow]$ & 0.1391 & 0.5371 & 4.2728 & 0.0193 \\
      1.0000 & $[\rightarrow\rightarrow\rightarrow]$ & 0.0001 & 0.0083 & 0.6206 & 1.6235 \\
      1.2033 & $[\downarrow \cdot \uparrow]$ & 0.0000 & 0.0031 & 0.4600 & 1.2794 \\
      1.5646 & $[\uparrow \uparrow \uparrow]$ & 0.0000 & 0.0005 & 0.2866 & 0.8937 \\
      1.7321 & $[\rightarrow \cdot \leftarrow]$ & 0.0000 & 0.0002 & 0.2341 & 0.7715 \\
      2.4083 & $[\rightarrow \leftarrow \rightarrow]$ & 0.0000 & 0.0000 & 0.1100 & 0.4594 \\
\hline
\hline
\multicolumn{6}{|c|}{$g=-0.005$}\\
\hline
 0.1593 & $[\Downarrow \Uparrow \Downarrow] \;{\scriptscriptstyle +}\; [\rightarrow \cdot \leftarrow]$ & 0.2974 & 0.9187 & 6.3014 & 13.0844 \\
  1.0000 & $ [\rightarrow\rightarrow\rightarrow]$ & 0.0001 & 0.0083 & 0.6206 & 1.6235 \\
  1.1674 & $ [\Downarrow \cdot \Uparrow] \;{\scriptscriptstyle +}\; [\rightarrow \leftarrow \rightarrow] $ & 0.0000 & 0.0037 & 0.4839 & 1.3313 \\
  1.5453 & $ [\downarrow \downarrow \downarrow] $ & 0.0000 & 0.0006 & 0.2935 & 0.9096 \\
  1.7478 & $[\Rightarrow \cdot \Leftarrow] \;{\scriptscriptstyle +}\; [\uparrow\downarrow\uparrow] $ & 0.0000 & 0.0002 & 0.2297 & 0.7612 \\
  2.3922 & $[\Rightarrow \Leftarrow \Rightarrow] \;{\scriptscriptstyle +}\; [\uparrow \cdot \downarrow] $ & 0.0000 & 0.0000 & 0.1119 & 0.4646 \\
\hline
\hline
\multicolumn{6}{|c|}{$g=-0.1$}\\
\hline
 0.6102 & $[\Downarrow \Uparrow \Downarrow] \; {\scriptscriptstyle  +} \; [\rightarrow \cdot \leftarrow]$ & 0.0029 & 0.0565 & 1.2559 & 2.9391 \\
 0.8873 & $ [\Downarrow \cdot \Uparrow] \;{\scriptscriptstyle +}\; [\rightarrow \leftarrow \rightarrow] $ & 0.0002 & 0.0143 & 0.7443 & 1.8837 \\
 1.0000 & $ [\rightarrow\rightarrow\rightarrow]$ & 0.0001 & 0.0083 & 0.6206 & 1.6235 \\
 1.4697 & $ [\downarrow \downarrow \downarrow] $ & 0.0000 & 0.0009 & 0.3227 & 0.9760 \\
 1.9313 & $[\Rightarrow \cdot \Leftarrow] \;{\scriptscriptstyle +}\; [\uparrow\downarrow\uparrow] $ & 0.0000 & 0.0001 & 0.1857 & 0.6550 \\
 2.1425 & $[\Rightarrow \Leftarrow \Rightarrow] \;{\scriptscriptstyle +}\; [\uparrow \cdot \downarrow] $ & 0.0000 & 0.0000 & 0.1467 & 0.5567 \\
\hline
\end{tabular} 
\caption{\label{Table:2}Mean vibrational number for each normal mode of the different initial structures, determined by the choice of $g$. The corresponding temperatures are given in $\mu\si{\kelvin}$. In the second row, the vibrations of the ions for each mode are sketched, and are displayed as sums of the eigenmodes of the linear configuration. The modes of the zigzag chain are composed of two normal modes of the linear chain, one of which (denoted by the thicker arrows) has the main contribution.}
\end{table}

\begin{figure}[htbp]
  \centering
  \subfloat{\label{fig:thermalsignal:lin}\phantom{(a)}\includegraphics[width=0.42\textwidth]{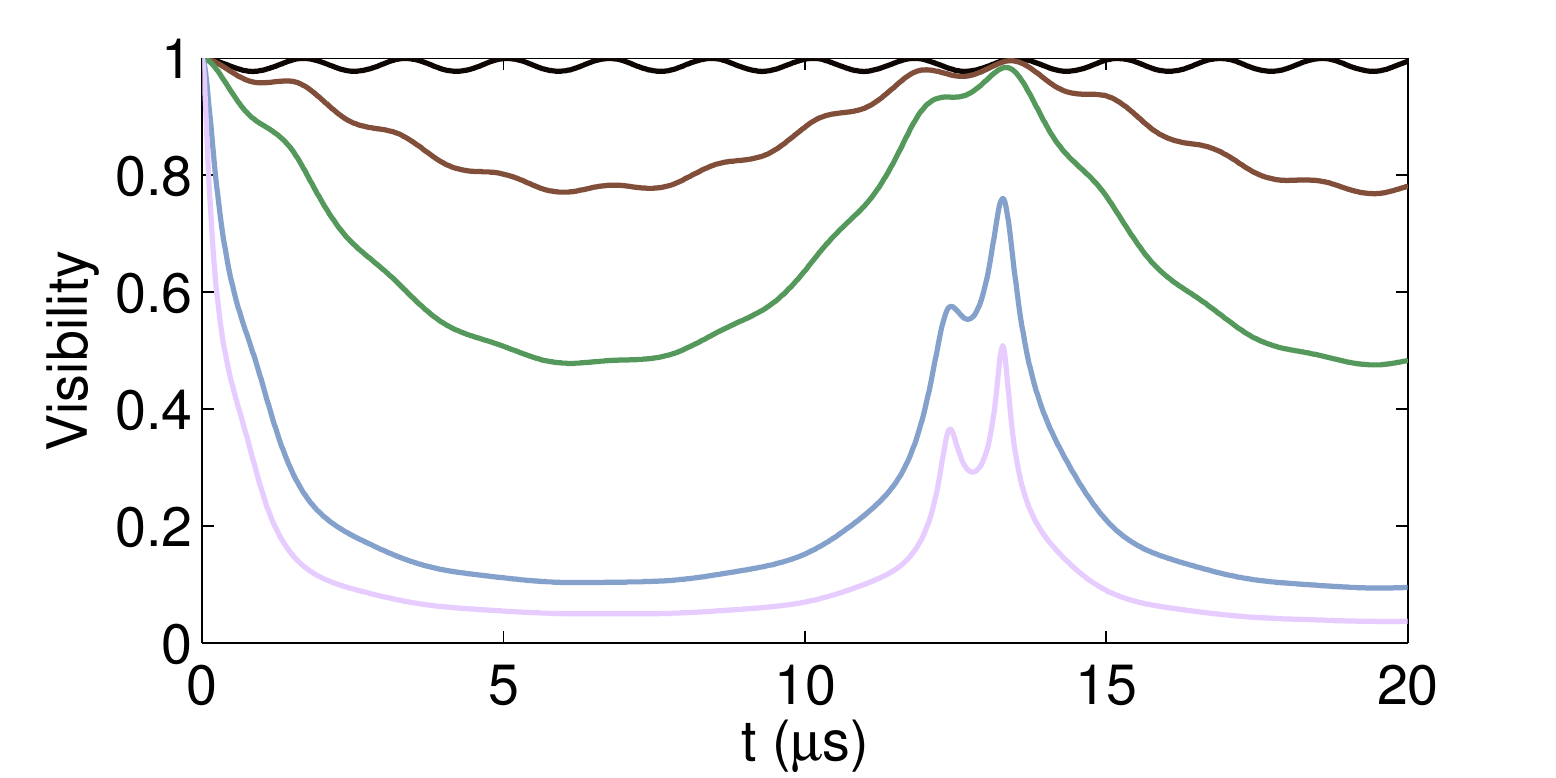}\put(-225,94){(a)}} 
  \\ \vspace{-.35cm}
  \subfloat{\label{fig:thermalsignal:sup}\phantom{(a)}\includegraphics[width=0.42\textwidth]{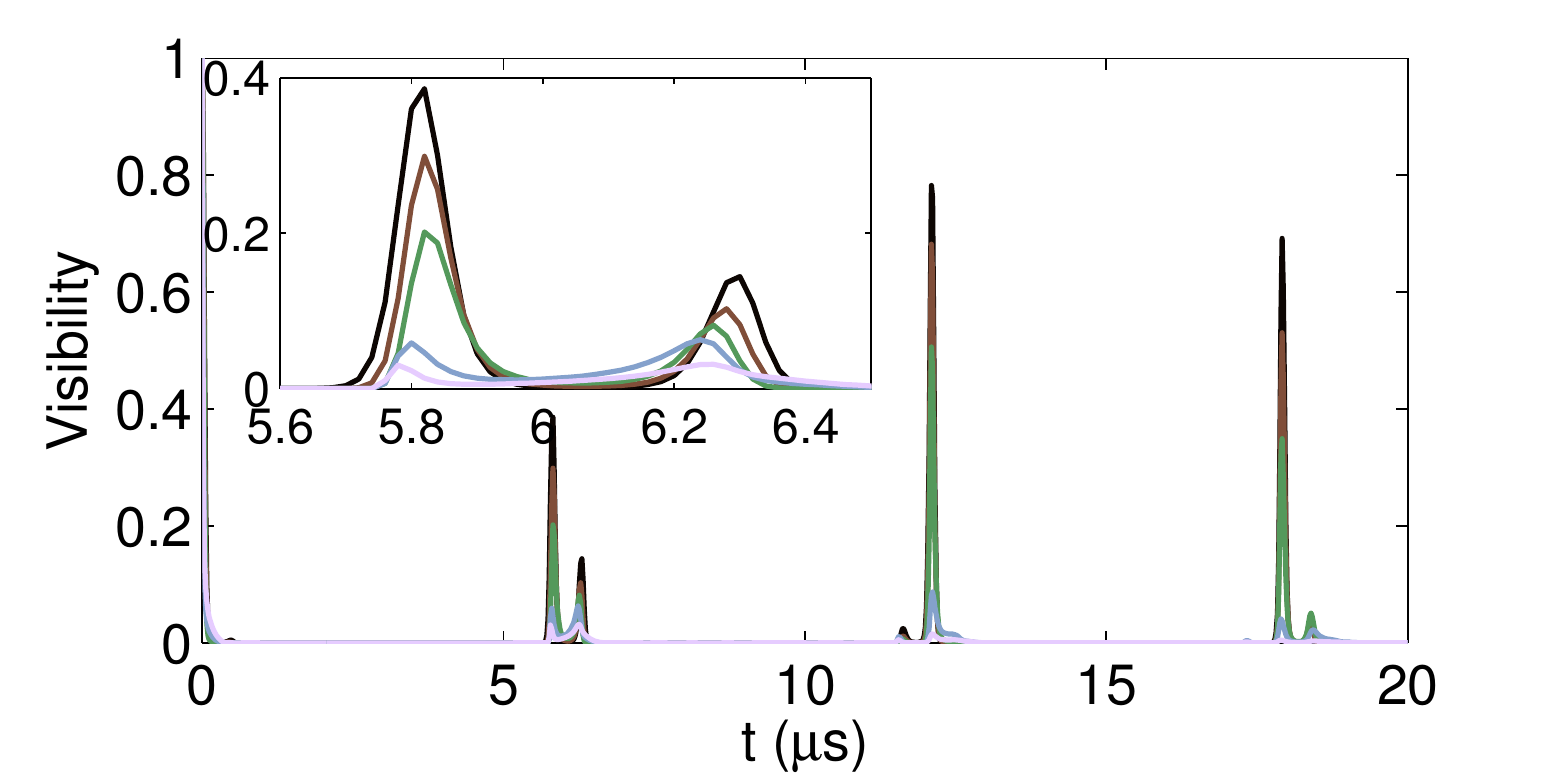}\put(-225,94){(b)}} \\ \vspace{-.35cm} 
  \subfloat{\label{fig:thermalsignal:zz}\phantom{(a)}\includegraphics[width=0.42\textwidth]{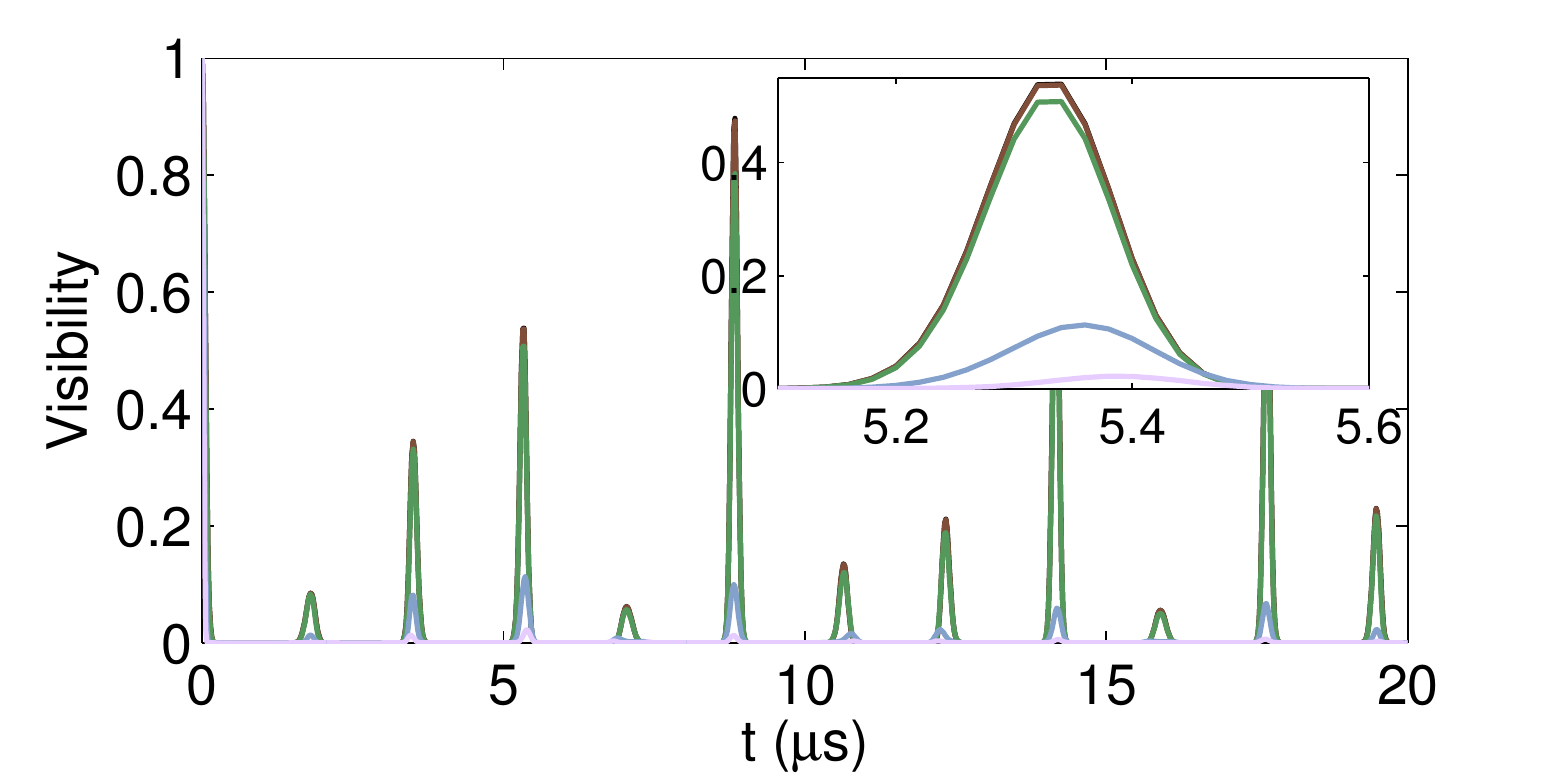}\put(-225,94){(c)}}
  \caption{(Color online) Visibility signal as a function of the time $t$ elapsed between the two Ramsey pulses. The signal is evaluated for temperatures $0\,\mu\si{\kelvin} $ (black line), $5\,\mu\si{\kelvin}$ (brown/dark gray line), $10\,\mu\si{\kelvin}$ (green/medium gray line), $50\,\mu\si{\kelvin}$ (blue/medium  light gray line) and  $100\,\mu\si{\kelvin}$ (light pink/light gray line). The parameters are  $\Delta = 0.025$ and (a)~$g=0.02$, (b)~$g=-0.005$, (c)~$g=-0.1$. The insets display a zoom of (b) the first double-peak and of~(c) the third peak.}
  \label{fig:thermalsignal}
\end{figure}
The corresponding visibility is displayed in Fig.~\ref{fig:thermalsignal} when the strength of the quench is $\Delta=0.025$. Panel~(a) displays the visibility at different temperatures for $g=0.02$, namely, when the initial and final state of the quench correspond to excitations of a linear structure. The visibility for $T=0$ is given by the black line and it oscillates between unity and a value above $0.95$. The oscillation is at the frequency of the zigzag eigenmode, which is excited by the quench~\cite{Baltrusch:2012}. As the temperature is increased the visibility decays, it still exhibits a modulation, which is markedly at a smaller frequency but at a larger amplitude than in the case at $T=0$. The corresponding maxima are at a time-scale which is independent of the temperature and exhibit a double-peak structure, which becomes evident at sufficiently large temperatures. Panel~(b) displays the visibility when the chain is initially a zigzag structure and the quench is performed across the linear-zigzag instability. The feature characterising the behaviour at $T=0$ is the rapid decay of the visibility to zero, and then the appearance of revivals the period of the soft mode. This feature is independent of the number of ions~\cite{Baltrusch:2012}. Increasing the temperature leads to a decrease of the amplitude of the revivals, as also visible in the inset: The amplitude significantly drops already at $T = 5\,\mu\si{\kelvin}$. Panel~(c) displays the visibility when the quench connects two different zigzag structures. Also in this case, at $T=0$ the visibility rapidly decays and then exhibits periodic revivals. Thermal effects lead to a decay of the amplitude of the revivals. Here, however, the signal is not significantly altered at temperatures as low as $T =  10\,\mu\si{\kelvin}$, as one can observe in the inset. 

\begin{figure}[bt]
  \centering
  \subfloat{\label{fig:thermalspectrum:strong:lin}\phantom{(a)}\includegraphics[width=0.4\textwidth]{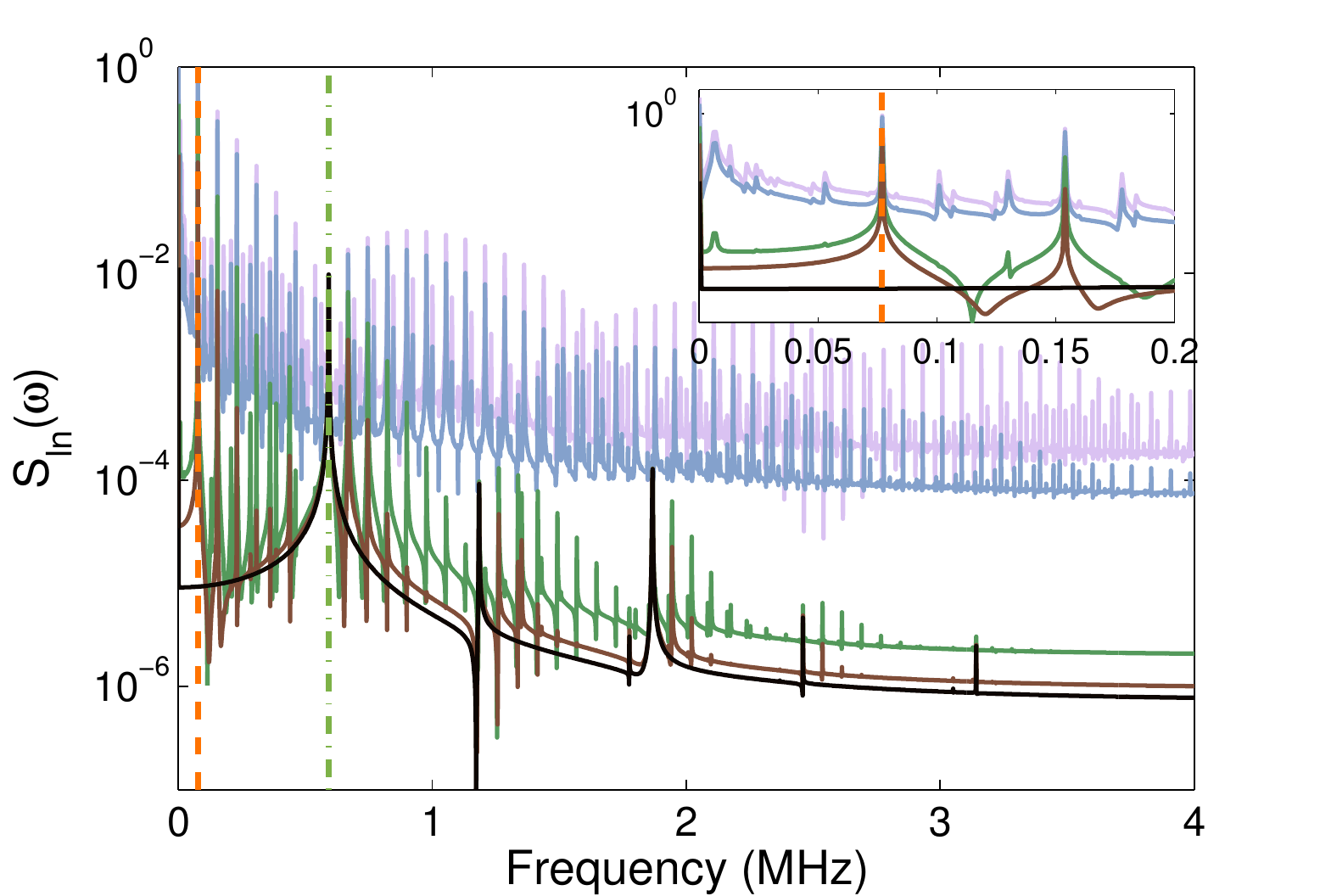}\put(-215,120){(a)}} 
  \\ \vspace{-.35cm}
  \subfloat{\label{fig:thermalspectrum:strong:trans}\phantom{(a)}\includegraphics[width=0.4\textwidth]{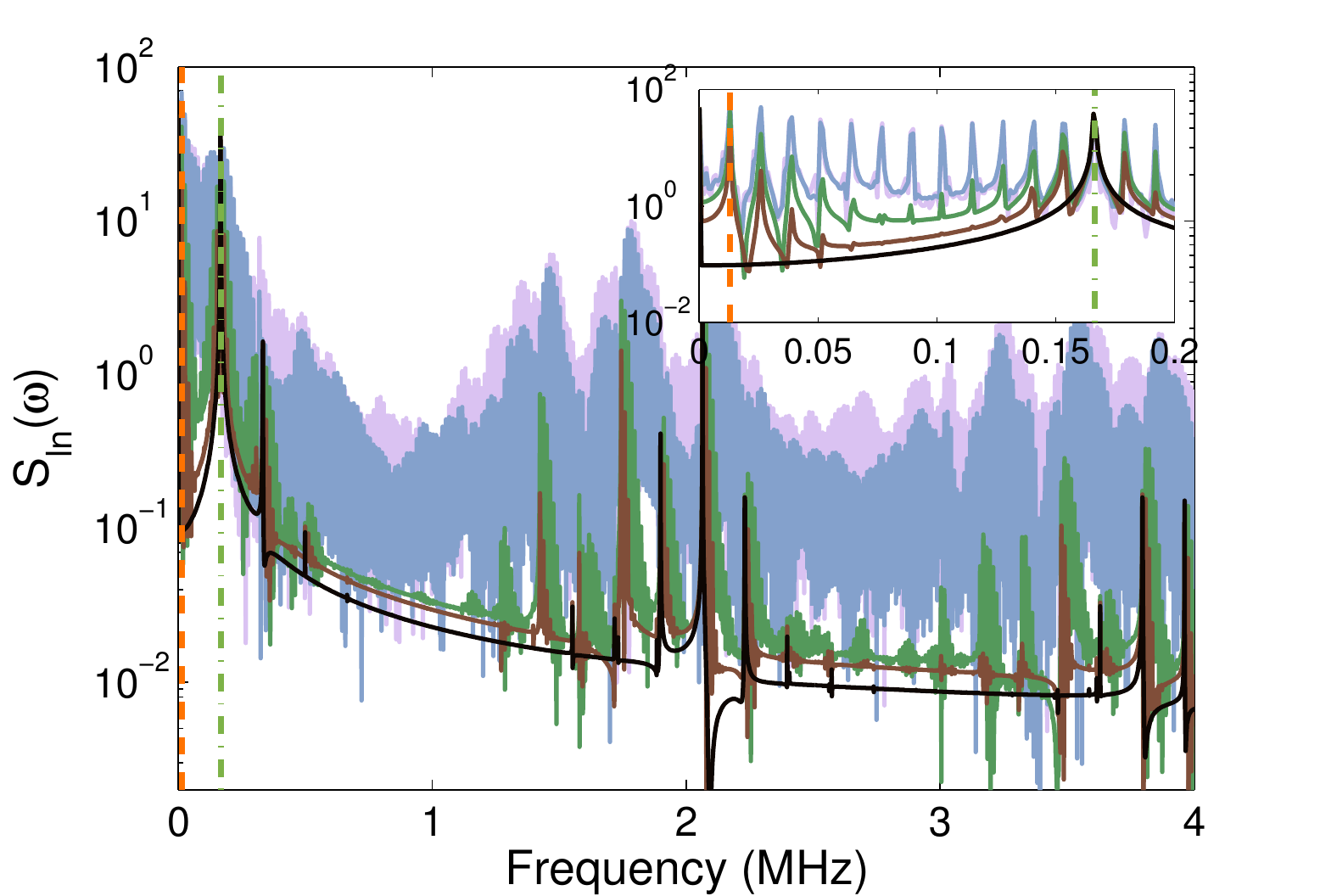}\put(-215,120){(b)}} 
  \\ \vspace{-.35cm}
  \subfloat{\label{fig:thermalspectrum:strong:zz}\phantom{(a)}\includegraphics[width=0.4\textwidth]{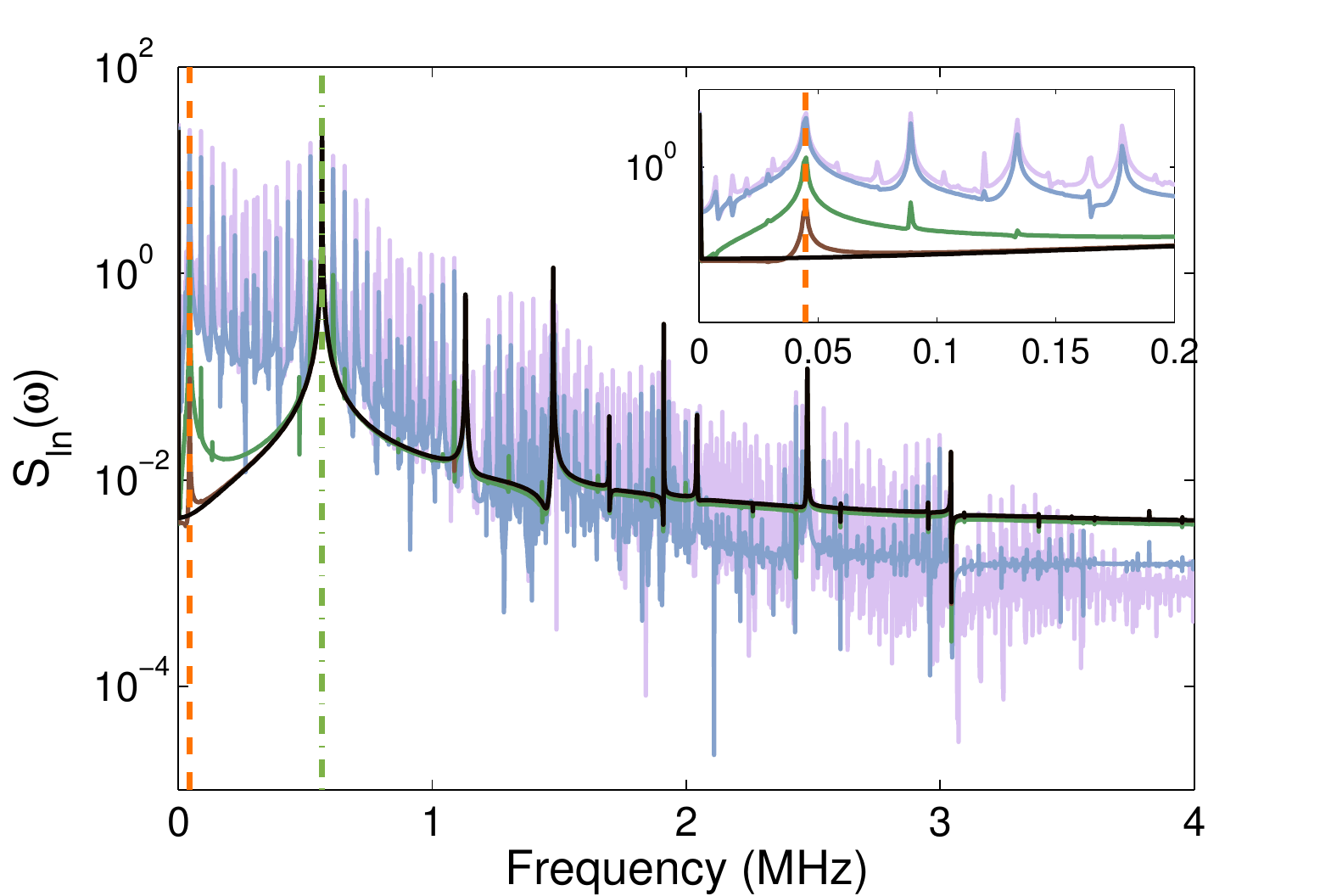}\put(-215,120){(c)}}
  \caption{\label{fig:thermalspectrum:strong}(Color online) Spectra for the logarithmic signals, Eq.~\eqref{F:log}, for the visibility curves in Fig.~\ref{fig:thermalsignal}. The vertical dash-dotted (green) lines show the zigzag eigenfrequency $\omega_1^e$ (in~(a) the line is at~$2\omega_1^e$). The dashed (orange) line shows the location of frequency $\omega_{\rm beat}=|\omega_1^e-\omega_1^g|$. The insets display a zoom of the low frequency part of the corresponding spectrum, highlighting the peak structure of the spectrum at multiples of~$\omega_{\rm beat}$.}
\end{figure}

These features can be better understood by analysing the spectrum of the signal. In particular, we choose to study the spectrum of the logarithmic visibility~\cite{Baltrusch:2012}, defined by
\begin{equation}
\label{F:log}
 {S}_{\ln}(\omega_n) = \frac{1}{T} \int_0^T \mathrm dt \; \ln [ \mathcal V(t) ] \; \mathrm{e}^{-{\mathrm{i}} \omega_n t}       \,.
\end{equation}
Figure~\ref{fig:thermalspectrum:strong} displays the spectra of the logarithmic visibility corresponding to the curves in Fig.~\ref{fig:thermalsignal}. Let us first recall the behaviour of the spectra at $T=0$. These exhibit well defined peaks at the frequency, or at multiples, of the soft mode. As the temperature increases the corresponding peaks are broadened. Moreover, additional peaks appear that are located at the beat frequency $\omega_{\rm beat}=|\omega_1^e-\omega_1^g|$ between the zigzag eigenmodes of the two structures, namely, the equilibrium structure when the ion is in the ground state and the one in which the ion is in the excited state. The appearance of a peak at this beating frequency is due to the fact that the corresponding mode in the initial configuration is thermally excited. The number of peaks increases with the temperature; they appear at multiples of $\omega_{\rm beat}$. This behaviour shows that the eigenmodes which most relevantly contribute to the overlap integral, and thus to the visibility, are the soft modes of the initial and excited structures, while the contribution of the other modes is marginal. Note that the soft modes are, for the parameters here considered, the ones which are at lowest frequency and significantly occupied, as one can see from Table~\ref{Table:2}.

\begin{figure}[tbp]
  \centering
  \subfloat{\label{fig:thermalsignal_comp:lin}\phantom{(a)}\includegraphics[width=0.4\textwidth]{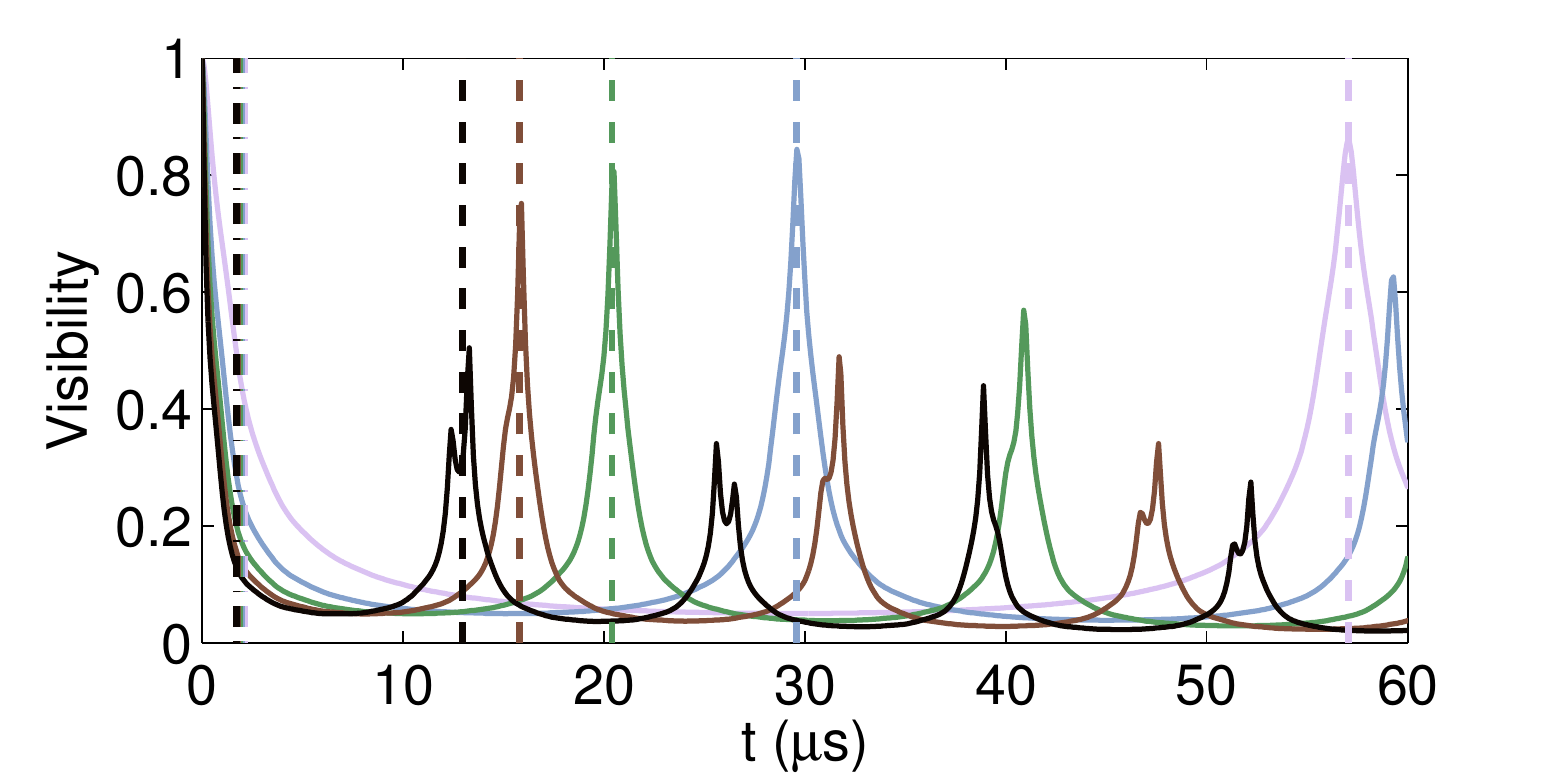}\put(-215,90){(a)}} 
  \\ \vspace{-.35cm}
  \subfloat{\label{fig:thermalsignal_comp:zz}\phantom{(a)}\includegraphics[width=0.4\textwidth]{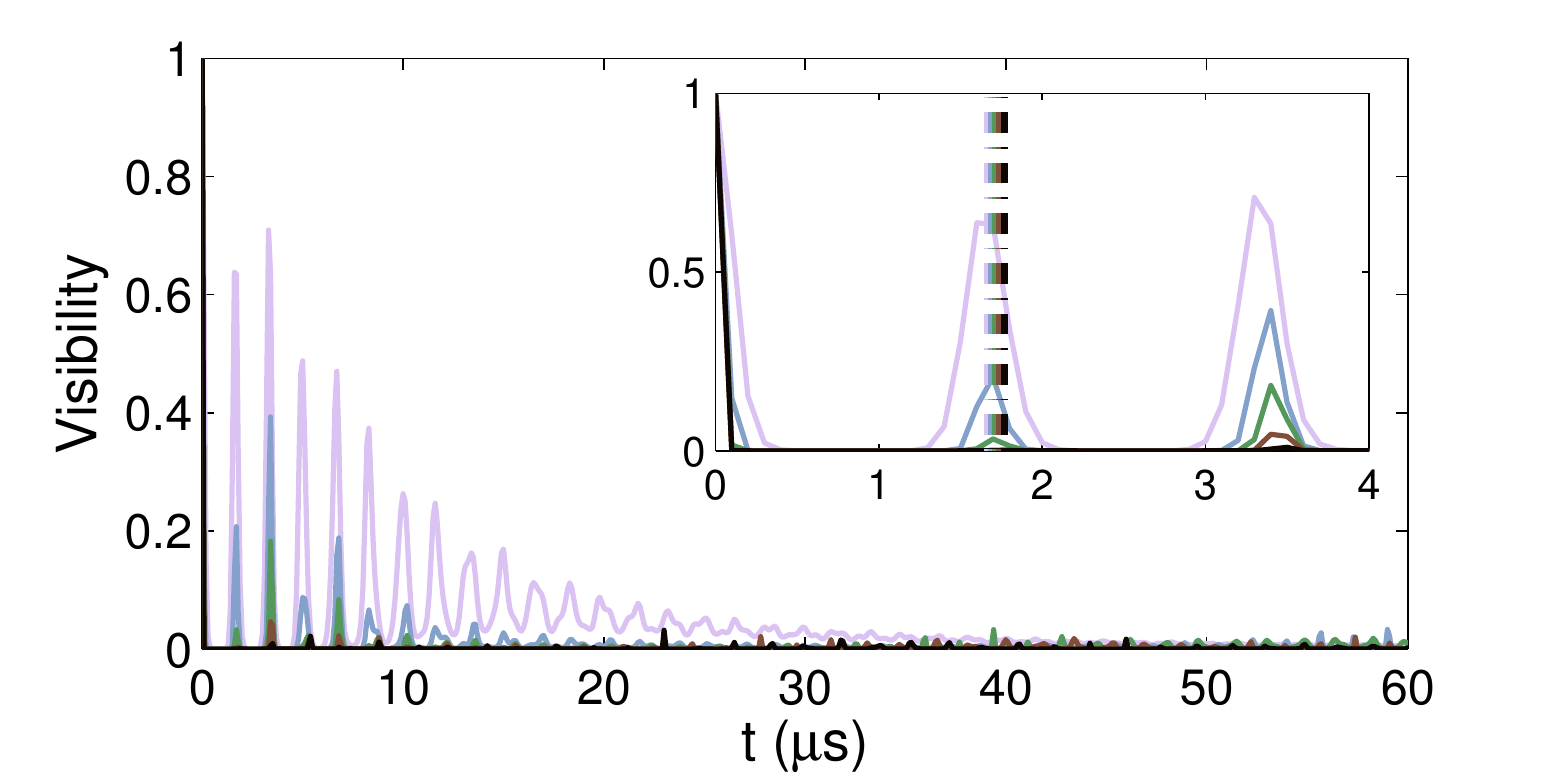}\put(-215,90){(b)}} 
  \caption{(Color online) Visibility as a function of the time $t$ elapsed between the two Ramsey pulses for $T=100\,\mu\si{\kelvin}$ and $\Delta=0.005$ (light pink), $\Delta=0.010$ (blue), $\Delta=0.015$ (green),  $\Delta=0.020$ (brown),  $\Delta=0.025$ (black line). In a scale of gray, as $\Delta$ increases the line becomes darker. The quench is performed for (a) $g=0.02$ and (b)~$g=-0.1$, such that the initial and quenched equilibrium strcutures are either both linear or zigzag chains.}
  \label{fig:thermalsignal_comp}
\end{figure}
We now compare the signals obtained for different strength $\Delta$ of the quench at finite temperature. Figure~\ref{fig:thermalsignal_comp} displays the visibility as a function of the elapsed time $t$ evaluated for different values of~$\Delta$ and when the chain is initially at temperature T=$100\,\mu\si{\kelvin}$. The case in which the two structures are linear is shown in panel~(a). Here, the peaks arising from initial thermal occupation wander to later times for smaller values of $\Delta$. This can be understood by recalling that these peaks are determined by the beating between the zigzag modes of the initial and of the quenched structure: For weaker quenches, the difference becomes smaller and the associated timescale, which is the period of the beating, correspondingly longer. Panel~(b) shows the behaviour when the initial and the quenched structures are both zigzag. Here, the visibility rapidly decays to zero, and then exhibits some revivals whose height also damps down to zero. This latter decay is slower for weaker quenches, i.e., for smaller values of $\Delta$. 

\begin{figure}[tbp]
  \centering
  \subfloat{\label{fig:thermalsignal_long_weak:lin}\phantom{(a)}\includegraphics[width=0.4\textwidth]{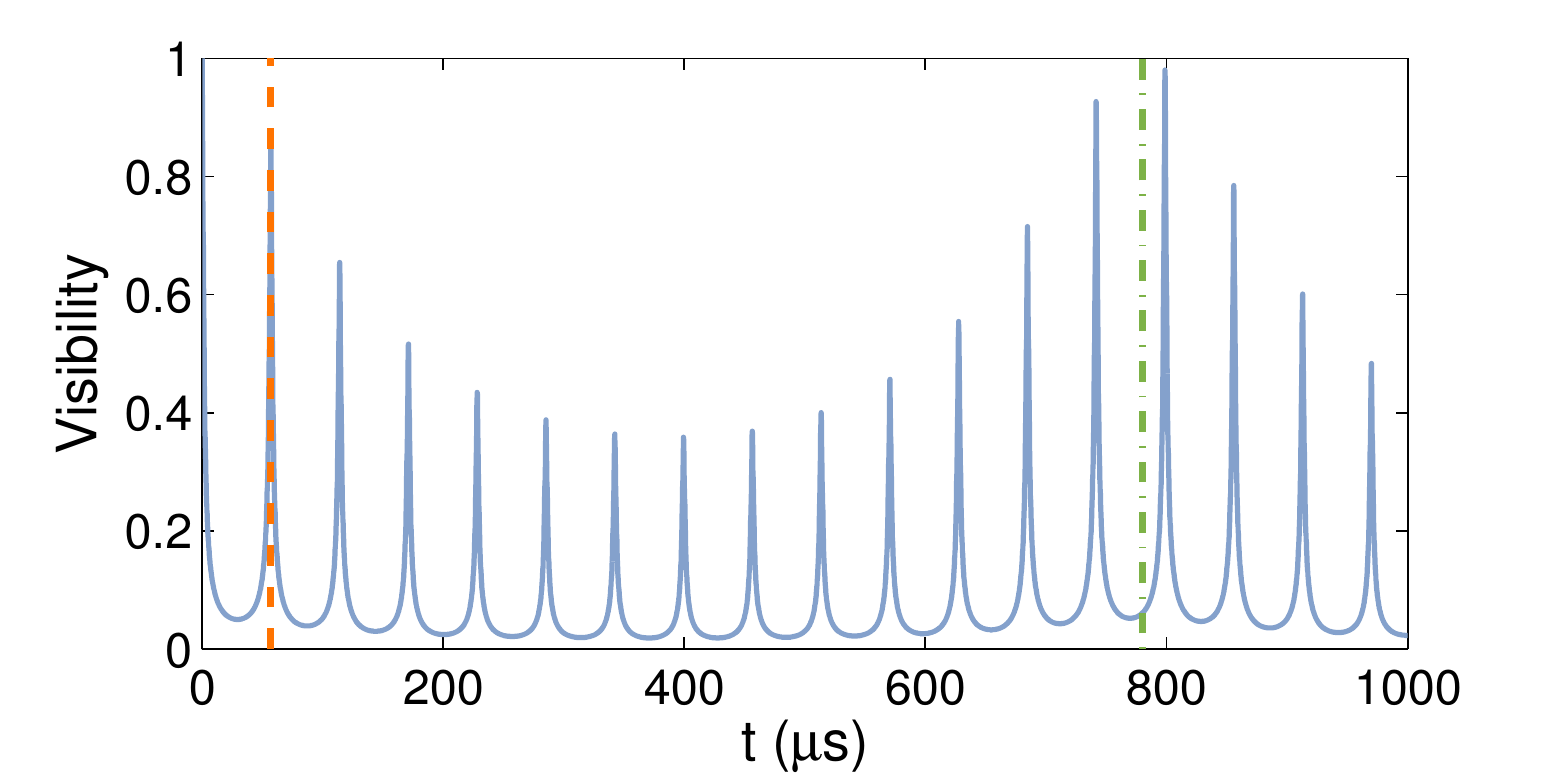}\put(-215,90){(a)}} 
  \\ \vspace{-.35cm}
  \subfloat{\label{fig:thermalsignal_long_weak:zz}\phantom{(a)}\includegraphics[width=0.4\textwidth]{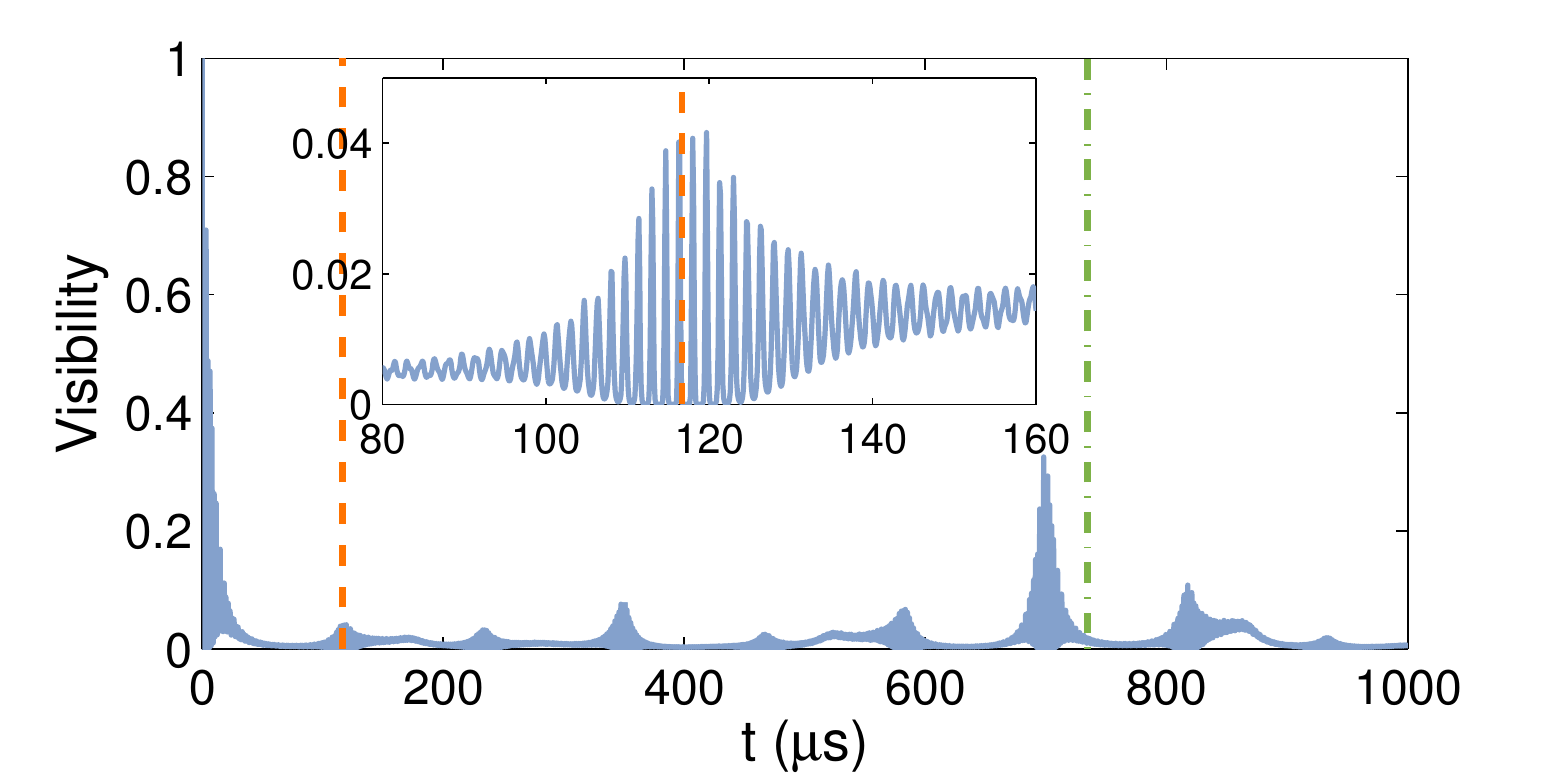}\put(-215,90){(b)}}
  \caption{(Color online) Visibility as a function of the elapsed time $t$ for $100\,\mu\si{\kelvin}$, $\Delta = 0.005$, and (a) $g=0.02$ and (b) $g=-0.1$. The vertical lines indicate the location of $\omega_{\rm beat}$ (dashed orange) and of the beating frequency between the transverse center-of-mass frequencies (dash-dotted green). The inset in (b) shows a zoom for the interval centered about~$\omega_{\rm beat}$.} 
\label{fig:thermalsignal_long_weak}
\end{figure}

The signal at $\Delta=0.005$ is singled out in Fig.~\ref{fig:thermalsignal_long_weak}, where it is plotted for a larger interval of elapsed times $t$. In Fig.~\ref{fig:thermalsignal_long_weak:lin}, where the initial and quenched structures are linear, the signal shows a slow modulation and a certain regularity. In Fig.~\ref{fig:thermalsignal_long_weak:zz}, where both structures are zigzag, the main peaks of the signal are less regularly distributed and exhibit a fast quasi-periodic modulation (see inset). In both situations one observes that at large times the visibility can be significantly above zero, showing that coherence persists over time scales of the order of milliseconds. 

\begin{figure}[tbp]
  \centering
  \subfloat{\label{fig:thermalspectrum:weak:lin}\phantom{(a)}\includegraphics[width=0.4\textwidth]{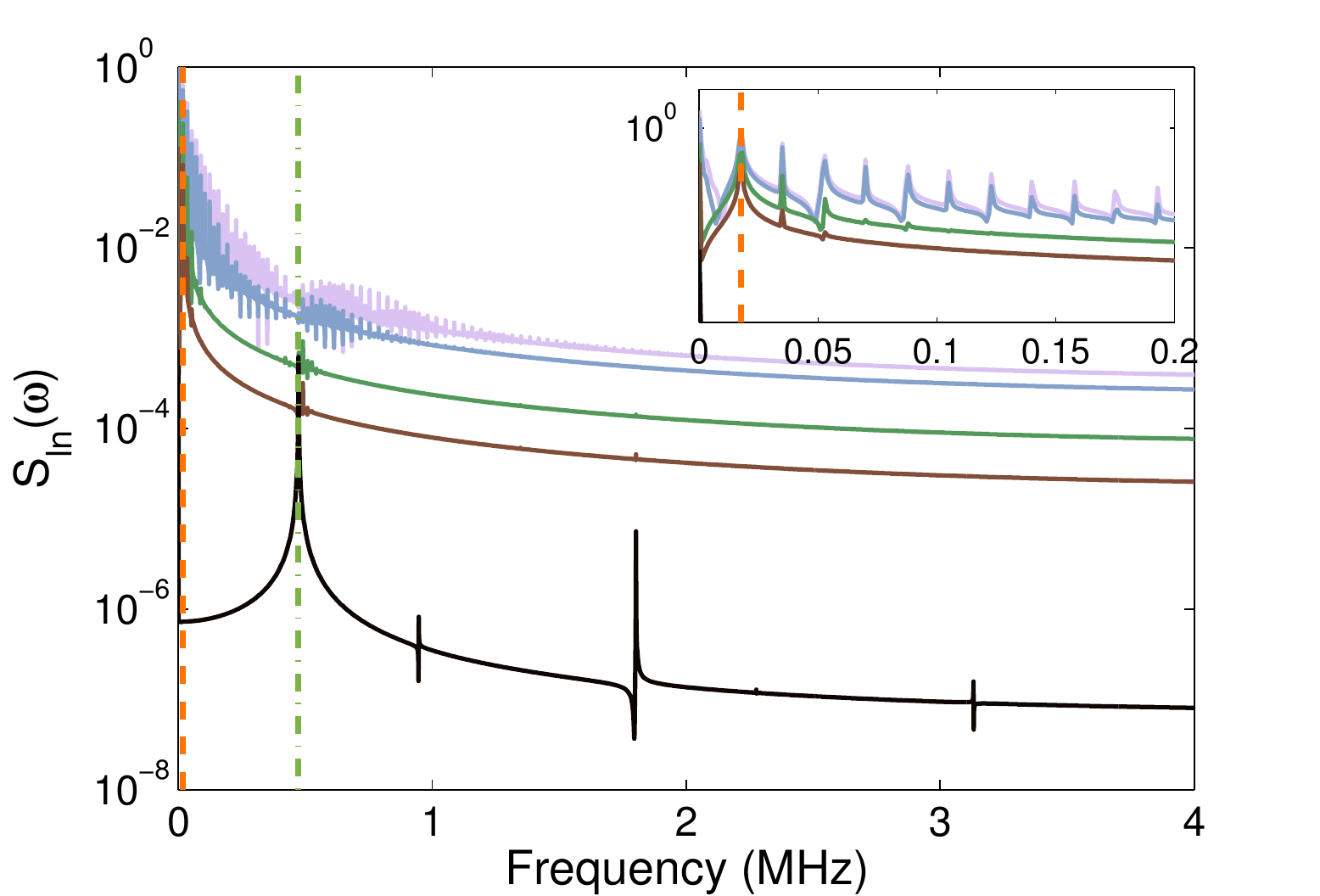}\put(-215,120){(a)}} 
  \\ \vspace{-.35cm}
  \subfloat{\label{fig:thermalspectrum:weak:zz}\phantom{(a)}\includegraphics[width=0.4\textwidth]{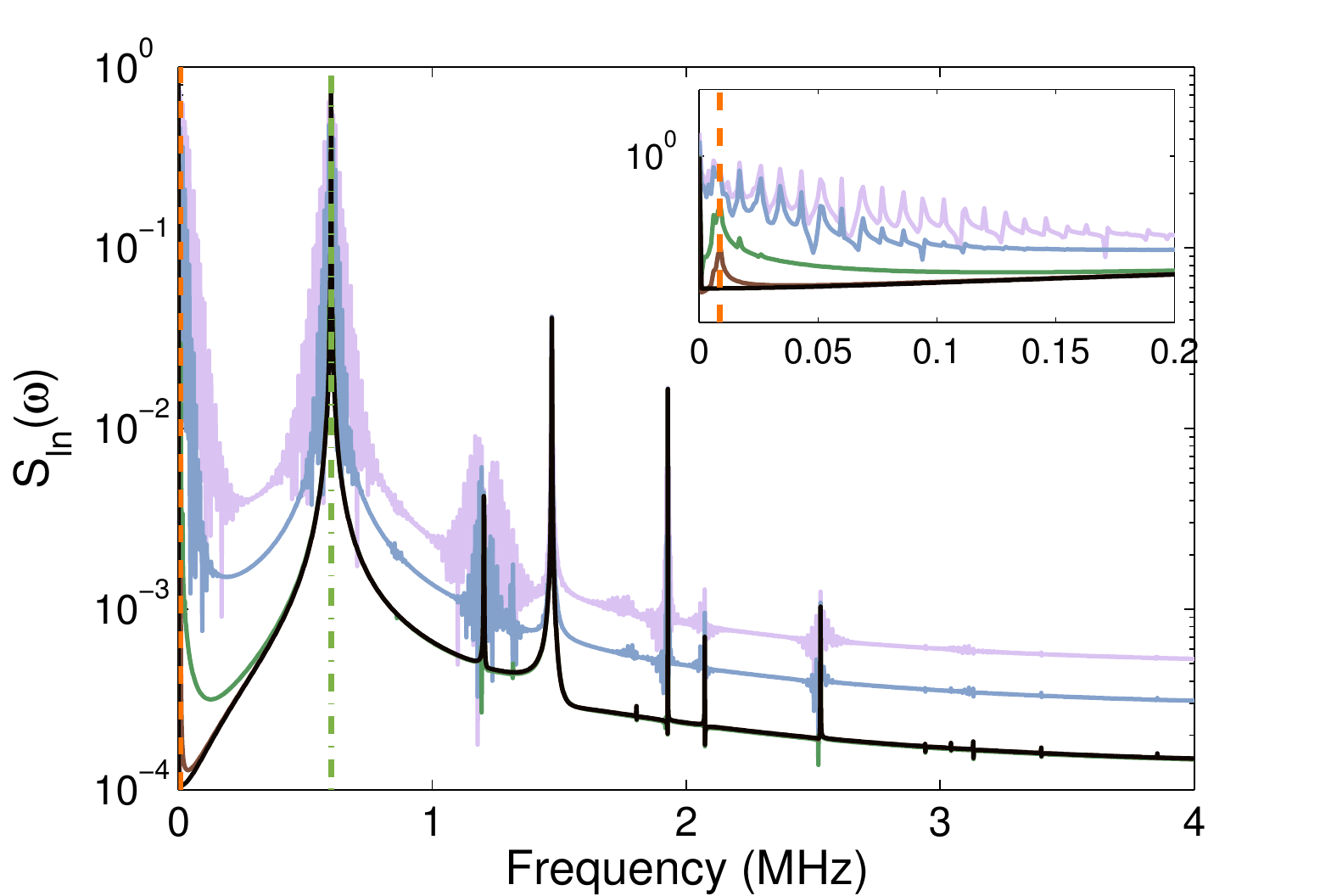}\put(-215,120){(b)}}
  \caption{(Color online) Spectrum of the logarithm of the visibility, Eq.~\eqref{F:log}, for $\Delta = 0.005$ and (a) $g=0.02$ (linear to linear) and (b) $g=-0.1$ (zigzag to zigzag). The curves correspond to the temperatures 
$T= 0$ (black line), $T=5\,\mu\si{\kelvin}$ (brown), $T=10\,\mu\si{\kelvin}$ (green), $T=50\,\mu\si{\kelvin}$ (blue) and $T=100\,\mu\si{\kelvin}$ (light pink), corresponding to a scale of grey from dark to light. The inset is a zoom in the low-frequency part. The vertical dash-dotted (green) line shows the location of the zigzag eigenfrequency in panel (b) and of  double the zigzag frequency in panel (a). The vertical dashed (orange) line shows the location of $\omega_{\rm beat}$.}
  \label{fig:thermalspectrum:weak}
\end{figure}

The relevant time scales associated with these features become evident by studying the spectrum of the logarithm of the visibility. Figure~\ref{fig:thermalspectrum:weak:lin} displays the spectrum corresponding to the signal in Fig.~\ref{fig:thermalsignal_long_weak:lin}. The additional curves, from top to bottom, correspond to decreasing values of the temperature. The black solid line reports the case $T=0$, which is here plotted for comparison. This curve displays clear peaks at values of the frequency corresponding to normal modes or to sum or difference of normal mode frequencies. The highest peak here corresponds to the frequency difference $|\omega^e_4 - \omega^g_4|$, which for the linear chain when all ions are in the ground state corresponds to center-of-mass oscillations in the transverse direction. This peak is still present at finite temperatures but becomes less prominent. On the other hand, at finite temperatures one observes the appearance of the peak at~$\omega_{\rm beat}$. Moreover, resonances at multiples of $\omega_{\rm beat}$ appear and their number increases with the temperature, as is evident by inspecting the inset. 

The spectrum in Fig.~\ref{fig:thermalspectrum:weak:zz} corresponds to the case in which the quench connects two zigzag configurations. Here, one observes that the peaks characterizing the spectrum at $T=0$ correspond to the frequencies of the normal modes; they are also present at finite temperatures, even though they become broader.  At finite $T$ a peak appears at $\omega_{\rm beat}$, while the number of harmonics increases with $T$, as visible in the inset.
\subsection{Discussion}
The evaluated visibility shows that finite temperatures lead to the appearance of various features, which emerge because of coherence between the two states created by the quench. The fact that the initial state is a statistical mixture leads to an overall decrease of the entanglement created by means of the quench. In particular, already at $T=100\,\mu\si{\kelvin}$ several features of the behaviour at zero temperature have disappeared. 

\begin{figure}[tbp]
  \centering
  \subfloat{\label{fig:thermalcompare_mode_cooling:lin}\phantom{(a)}\includegraphics[width=0.4\textwidth]{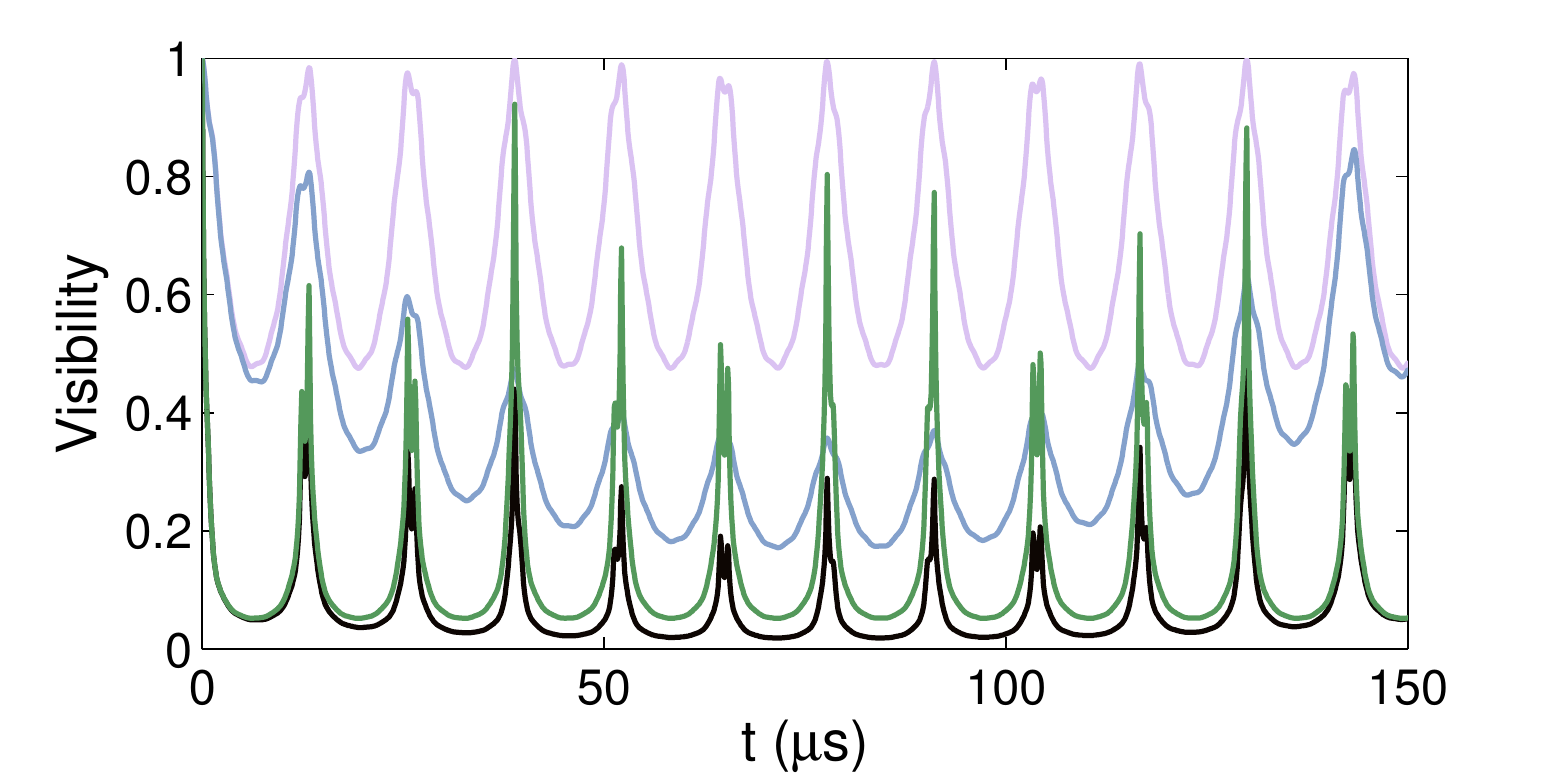}\put(-215,90){(a)}} 
  \\ \vspace{-.35cm}
  \subfloat{\label{fig:thermalcompare_mode_cooling:sup}\phantom{(a)}\includegraphics[width=0.4\textwidth]{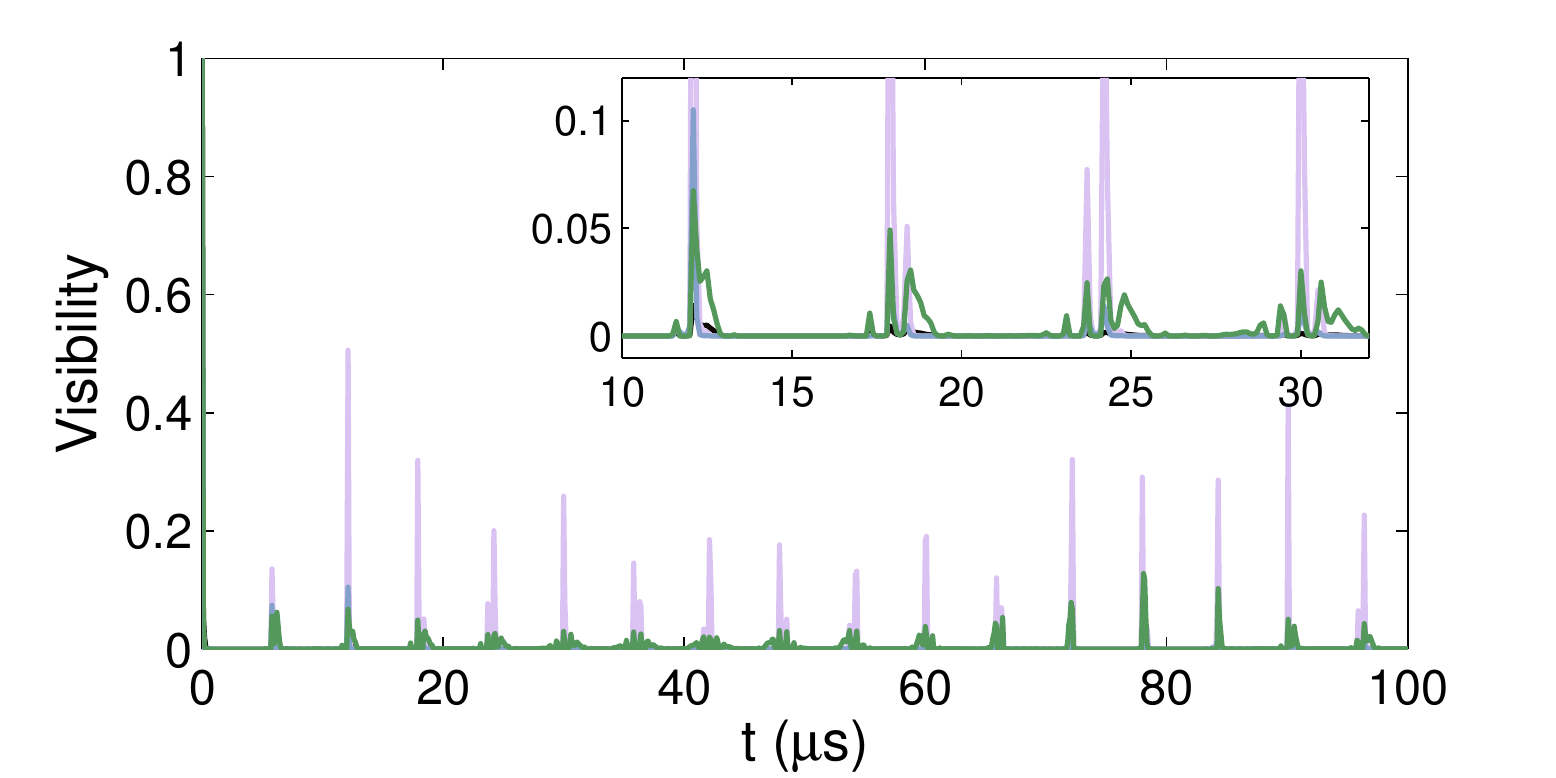}\put(-215,90){(b)}} 
  \\ \vspace{-.35cm}
  \subfloat{\label{fig:thermalcompare_mode_cooling:zz}\phantom{(a)}\includegraphics[width=0.4\textwidth]{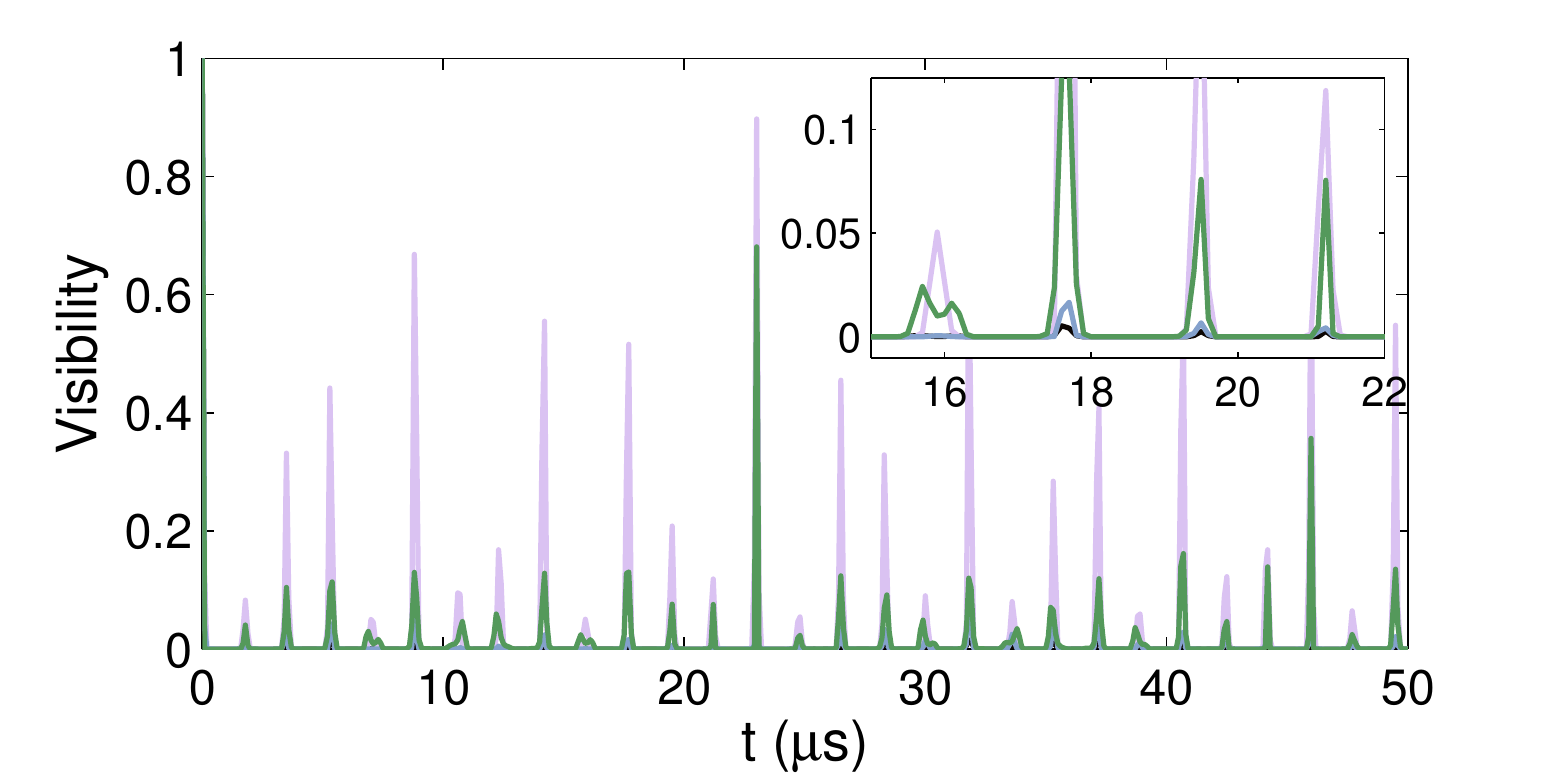}\put(-215,90){(c)}}
  \caption{(Color online) Visibility as a function of the elapsed time for $\Delta=0.025$ , and (a)~$g=0.02$, (b)~$g=-0.005$, (c)~$g=-0.1$. The pink and black lines correspond to the crystal initially at $T=10\,\mu\si{\kelvin}$ and  
 $100\,\mu\si{\kelvin}$, respectively.  The green (dark gray) line is for the zigzag mode at  $100\,\mu\si{\kelvin}$ and all other modes cooled at $10\,\mu\si{\kelvin}$, the blue (medium gray) line is for the zigzag mode cooled at $10\,\mu\si{\kelvin}$ and all other modes at $100\,\mu\si{\kelvin}$.}
  \label{fig:thermal_compare_mode_cooling}
\end{figure}

One signature of thermal excitation is the appearance of peaks at the harmonic of frequency $\omega_{\rm beat}$. These can be suppressed by cooling the zigzag mode, which is majorly excited by the quench, to a lower temperature. Figure~\ref{fig:thermal_compare_mode_cooling} displays the visibility signal as a function of the elapsed time when the zigzag mode has been cooled to  $10\,\mu\si{\kelvin}$ while the other modes are at $T=100\,\mu\si{\kelvin}$ (see the blue line). By comparing this behaviour with the visibility for the chain in the thermal state at $T=10\,\mu\si{\kelvin}$ and $100 \mu\si{\kelvin}$ we observe in (a) and (b) that for elapsed times of the order of tens of microseconds the visibility qualitatively reproduces the behaviour found by cooling all modes at $10\,\mu\si{\kelvin}$. Thus, for quenches connecting two linear structures (case a) or connecting a linear and a zigzag structure (case b) the initial excitation of the zigzag mode determines the visibility behaviour up to times of the order of $10\,\mu \si{\second}$. This is also confirmed when comparing with the opposite case, in which the whole chain has been cooled to $10\,\mu\si{\kelvin}$ except for the zigzag mode, whose vibrational excitations follow a thermal distribution corresponding to $100\,\mu\si{\kelvin}$. The visibility in this case behaves similarly to the one where all modes are at $T=100\,\mu\si{\kelvin}$. A different situation is found when the quench connects two zigzag structures and is displayed in panel~(c). Here, all eigenfrequencies contribute in determining the dynamics at low temperatures. 

We finally comment on the mechanical effect of light, which in certain configurations of laser pulses can also contribute to excite normal modes of the crystal. Its effect has been extensively studied in Ref.~\cite{DeChiara:2008} for different parameter regimes. Here, we just show how this may modify the signals by assuming that the states $\ket{g}$ and $\ket{e}$ are two hyperfine states of the ground state multiplet of $^9$Be$^+$ which are resonantly driven by two lasers of wave vectors ${\bf k_g}$ and ${\bf k_e}$ via a coherent Raman transition~\cite{Leibfried:2003}. The effective wave vector ${\bf k}$, which determines the mechanical momentum imparted by the light on the ion, is the difference between the wave vectors of the two beams and can thus range from zero, when the beams are copropagating, to twice the wave vector ${\bf k_g}$. We choose that the momentum imparted by the first pulse is equal and opposite to the one of the second pulse, ${\bf k} = {\bf k'}$, and is always along the transverse, $y$, direction. We consider three situations: (i) copropagating beams, i.e., ${\bf k}=0$; (ii) orthogonal beams, ${\bf k_g}\cdot {\bf k_e}=0$ (with the resulting wave vector along the $y$ axis), and (iii) counterpropagating beams along $y$, ${\bf k}=2{\bf k}_g$. In order to single out the effect of the photon recoil, we assume that the crystal is at temperature $T=0$. 

\begin{figure}[tbp]
  \centering
  \subfloat{\label{fig:photon_co_lin}\phantom{(a)}\includegraphics[width=0.4\textwidth]{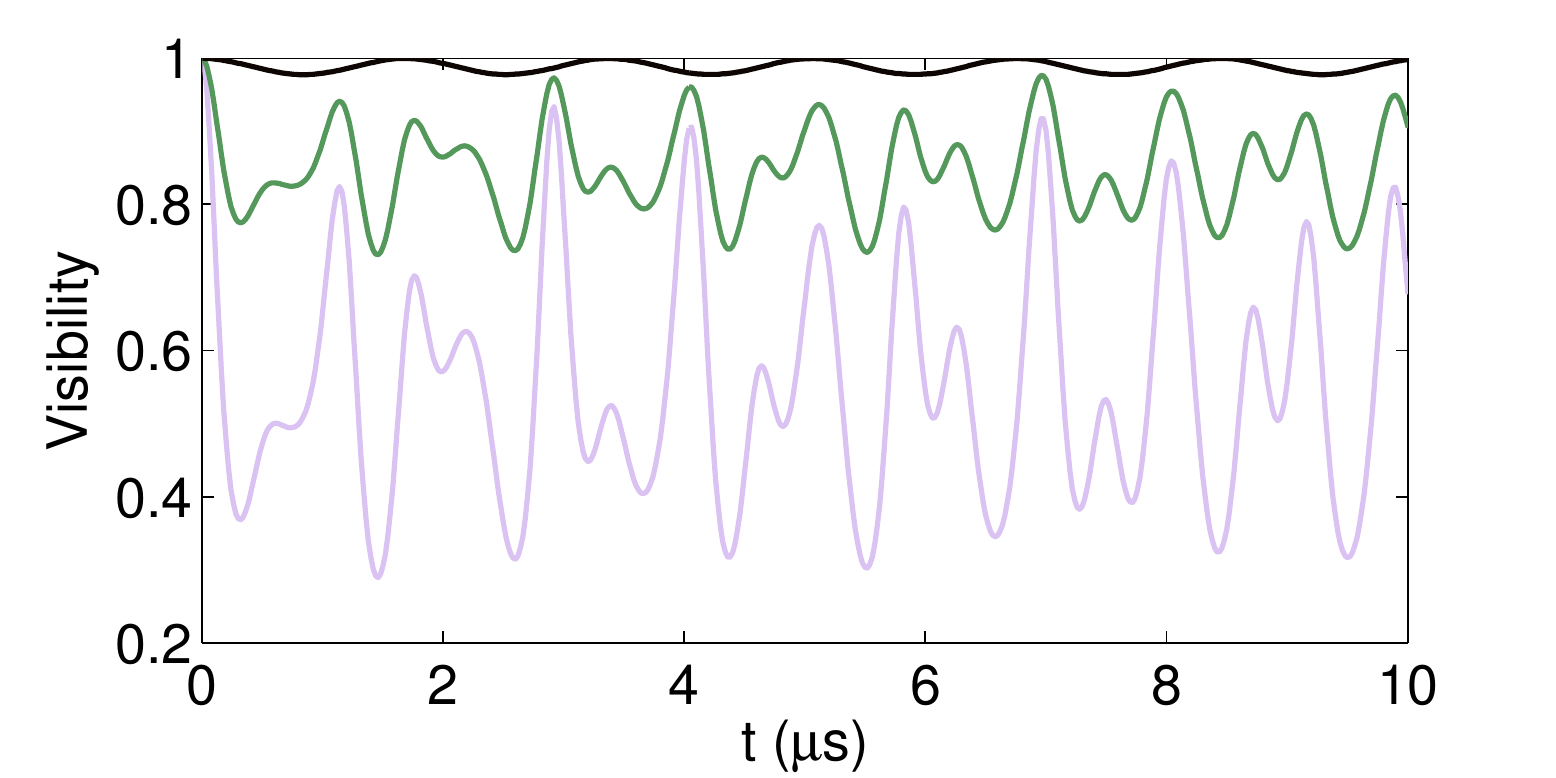}\put(-225,90){(a)}}
  \\ \hspace{-.3cm}
  \subfloat{\label{fig:photon_co_sup}\phantom{(a)}\includegraphics[width=0.4\textwidth]{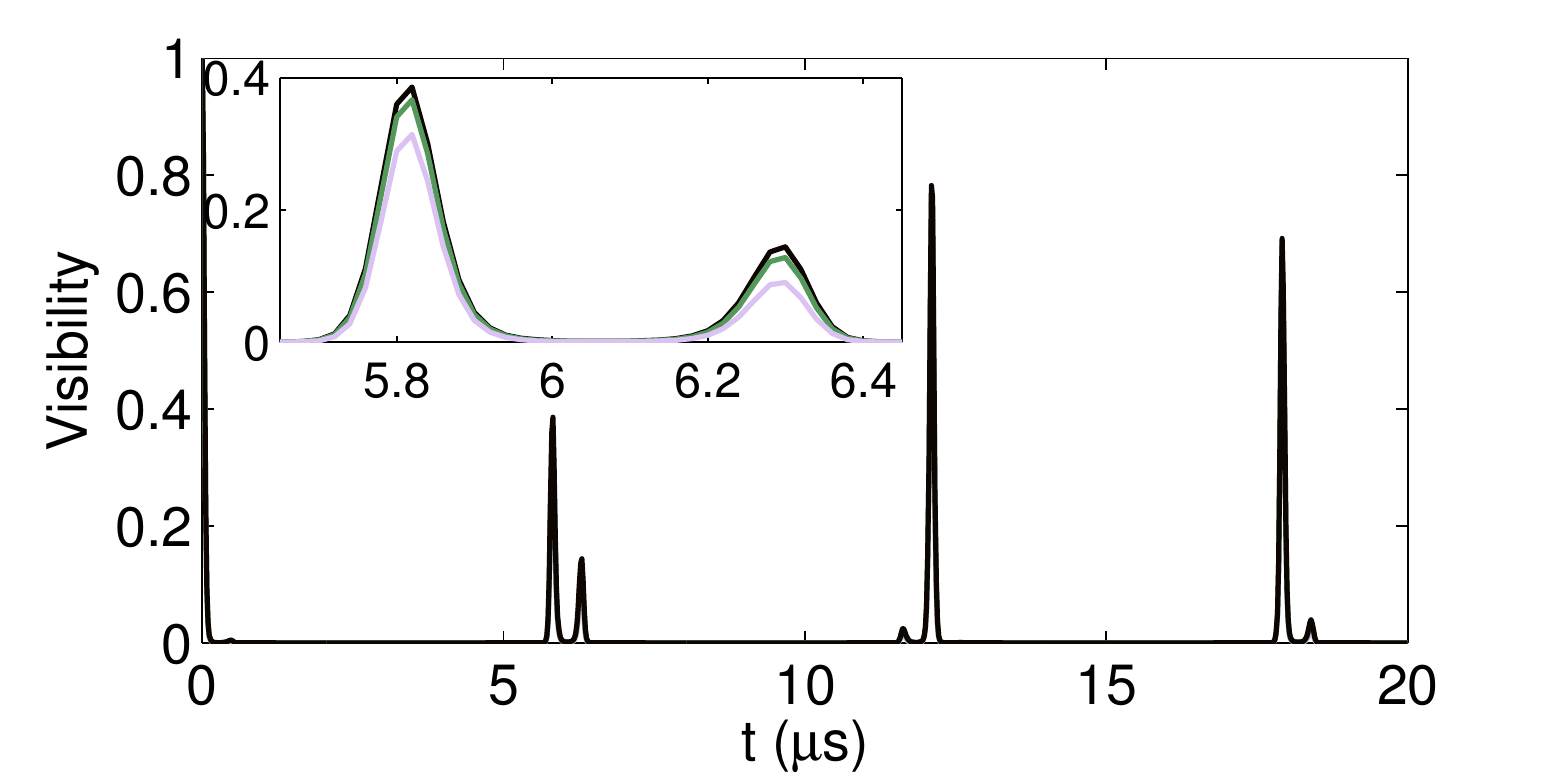}\put(-225,90){(b)}}
  \\ \hspace{-.3cm}
  \subfloat{\label{fig:photon_co_zz}\phantom{(a)}\includegraphics[width=0.4\textwidth]{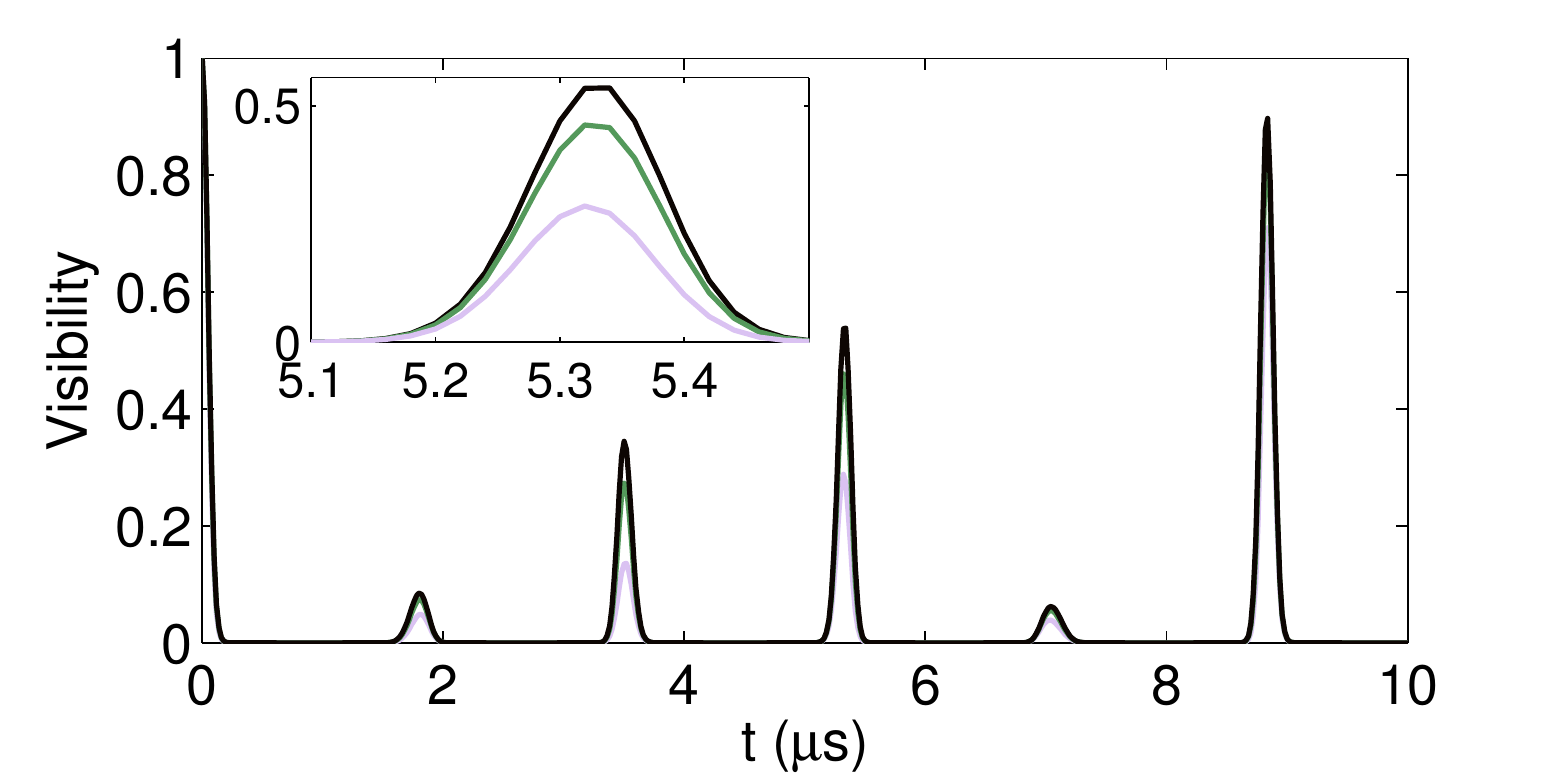}\put(-225,90){(c)}} 

  \caption{(Color online) Visibility as a function of the time $t$ elapsed between the Ramsey pulses for three ${}^9\mathrm{Be}^+$ ions at $T=0 \mathrm{K}$. The parameters are $\Delta = 0.025$ and (a)~$g=0.02$, (b)~$g=-0.005$, (c)~$g=-0.1$. A Raman transition of the central ion is driven by two laser beams at $313$nm which are copropagating (black line), orthogonal (green), or counterpropagating (pink line). In a scale of gray the three curves range from black to light gray. In panels~(b,c), the curves practically lie on top of each other. }
  \label{fig:photonrecoil}
\end{figure}

Figure~\ref{fig:photon_co_lin} displays the visibility as a function of the elapsed time when the quench connects two linear structures. The signal experiences a visible change due to the photon recoil, which is about the same magnitude as the variation of the signal when there is no photon recoil. In particular, one observes the appearance of other frequencies, which are due to the excitation of other normal modes by the pulse. 

The case in which the quench connects two zigzag structures is shown in Fig.~\ref{fig:photon_co_zz}. The curves lie on top of each other. The effect of the photon recoil is here insignificant on the scale of the variation of the signal which is originated by the quench. A similar behaviour is encountered when the quench is across the linear-zigzag transition, as visible in Fig.~\ref{fig:photon_co_sup}, where the excitation due to the quench dominates over the mechanical effect. This behaviour is due to the chosen parameters. Smaller quenches give rise to signals where the mechanical effects become more visible. 

\section{Conclusions}
\label{sec:5}
The dynamical properties of an ion crystal after a quench have been theoretically determined, when the crystal vibrations are in a thermal state and the quench is performed by creating coherent superpositions of motional states close to and across the linear-zigzag structural transition. These dynamics are partly revealed by performing Ramsey interferometry on one ion of the chain. The behaviour of the visibility as a function of the temperature has been characterised. 

This analysis shows that the dynamics we predict could be experimentally observed in small ion chains. The calculations we performed took parameters which are experimentally accessible and show that already at the temperatures typically achieved by means of Doppler cooling~\cite{Eschner:2003}, it is possible to measure the features we identify and which are related to the existence of mesoscopic quantum coherence between the structures. A prerequisite is that unitary evolution is warranted over time scales of the order of hundreds of microseconds, which is frequently fulfilled in state-of-the-art experiments \cite{Leibfried:2003,HaeffnerRoosBlattPhysRep2008,Schaetz:2012a}. Entanglement generated by the quench can be increased if the chain has been previously cooled to ultralow temperatures by means of sideband cooling~\cite{Eschner:2003,Leibfried:2003} or EIT cooling~\cite{EIT:2012} of the modes of the chain. 

Another important finding is that the dynamics excited by the quench is mostly dominated by the zigzag mode of the linear chain, which is the mode driving the linear-zigzag instability. A check on the Ramsey signal suggests that, after Doppler precooling, it would be sufficient to ground-state cool the zigzag mode, for instance by means of sideband cooling, in order to qualitatively reproduce the behaviour of the visibility found when the chain is at $T=0$. 

This study shows that these dynamics can be observed in existing experimental setups. Moreover, our formalism can be directly applied to larger chains and it can be extended by taking into account the normal modes including the effects of the micromotion~\cite{Landa:2012a,Landa:2012b}. The analysis in this work provides the basis for investigations on the onset of thermalization in closed quantum systems~\cite{MukherjeePRB2012,JacobsonPRA2011}.

\section*{Acknowledgments}
The authors acknowledge discussions with Christoph Wunderlich, Tommaso Calarco, Gabriele De Chiara, and Shmuel Fishman and support by the European Commission (Integrating Project ``AQUTE'', STREP ``PICC'', COST action ``IOTA''), the Spanish Ministery of Science (EUROQUAM ``CMMC'', Consolider Ingenio 2010), the German-Israeli Foundation, the Alexander von Humboldt and the German Research Foundations.  
\appendix
\section{Calculation of the Overlap Integral}
  \label{app:overlapintegral}
The integral we have to calculate is given by
\begin{align*}
  \mathfrak I_{\alpha}(\lambda^g) =&  \int \frac{\mathrm d^{6N} \alpha}{\uppi^{3N}} 
    \braketIX{e}{\alpha}{\theta}{e}
    \braketIX{e}{\theta'}{\alpha(t)}{e} \overbrace{\matrixElementSub{ \alpha}{e}{\mathrm{e}^{\mathrm{A}(\theta)}}{ \alpha}{e}}^{\langle \mathrm{e}^{\mathrm{A}(\theta)} \rangle_{\alpha} 
}\nonumber\\
    &\hphantom{\int \frac{\mathrm d^{6N} \alpha}{\uppi^{3N}} } \times     \underbrace{\matrixElementSub{ \alpha(t)}{e}{\mathrm{e}^{\mathrm{A}^\dagger(\theta')}}{ \alpha(t)}{e}}_{\langle \mathrm{e}^{\mathrm{A}^\dagger(\theta')}\rangle_{\alpha(t)}} \,.
\end{align*}
Evaluating all expressions, we get
\begin{align*}
  \langle \mathrm{e}^{\mathrm{A}(\theta)} \rangle_{\alpha} 
    &= \exp \Bigl\{\frac{1}{2} \sum_{jk} A_{jk} (\alpha_j^* - \theta_j^*)(\alpha_k^* -\theta_k^*)\Bigr\} \,, \\
  \langle \mathrm{e}^{\mathrm{A}^\dagger(\theta')}\rangle_{\alpha(t)}
    &= \exp \Bigl\{\frac{1}{2} \sum_{jk} A_{jk} (\alpha_j(t) - \theta'_j) (\alpha_k (t) - \theta'_k )\Bigr\} \,,
\end{align*}
and
\begin{align*}
 \braketIX{e}{\alpha}{\theta}{e} &= \exp \Bigl\{ \sum_j  \Bigl[ -\frac{\modulus{\alpha_j}^2}{2} - \frac{\modulus{\theta_j}^2}{2} + \alpha_j^* \theta_j  \Bigr] \Bigr\} \,, \\
 \braketIX{e}{\theta'}{\alpha(t)}{e} &= \exp \Bigl\{ \sum_j  \Bigl[ -\frac{\modulus{\theta'_j}^2}{2} - \frac{\modulus{\alpha_j(t)}^2}{2} + \theta'_j{}^* \alpha_j(t)  \Bigr] \Bigr\} \,.
\end{align*}

Merging all terms into a single exponential and sorting them by their orders at once, the exponent reads
 \begin{multline*}
  \frac{1}{2} \sum_{jk} 
  \begin{pmatrix}
    \alpha_j \\ \alpha_j^*
    \end{pmatrix}^{T}
  \begin{pmatrix}
    \widetilde{A}_{jk} & -\delta_{jk} \\ -\delta_{jk} & A_{jk} 
  \end{pmatrix}
  \begin{pmatrix}
    \alpha_k \\ \alpha_k^*
  \end{pmatrix} \\
 - \sum_j S_{j}[\theta] \alpha_j^* - \sum_j S_{j}^*[\theta'] \mathrm{e}^{-\mathrm{i}\omega^e_j t} \alpha_j +  G^*(\theta') + G(\theta) \,,
\end{multline*}
with the definitions of
\begin{align}
   \widetilde{A}_{jk} &=  A_{jk} \mathrm{e}^{-\mathrm{i} (\omega^e_j+\omega^e_k) t} \,, & S_j[\beta] &= \sum_{k} A_{jk} \beta_k^* - \beta_j \,,
\end{align}
and
\begin{equation}
  G(\beta) =   \sum_{jk}  \frac{A_{jk}}{2} \beta_j^*\beta_k^*   - \sum_{j} \frac{|\beta_j|^2}{2} \,.
\end{equation}

We now express the integration variables by their real and imaginary parts, $\alpha_j = u_j + i v_j$ and $\alpha_j^* = u_j - \mathrm{i} v_j$. 
The quadratic term is rewritten as
\begin{equation}
 - \sum_{jk}
 \begin{pmatrix}
  u_j \\ v_j 
 \end{pmatrix}^T
\begin{pmatrix}
 \delta_{jk} - A_{jk}^+ & \phantom{\delta_{jk}}-\mathrm{i} A_{jk}^- \\ \phantom{\delta_{jk}}-\mathrm{i} A_{jk}^- & \delta_{jk}+ A_{jk}^+
\end{pmatrix}
\begin{pmatrix}
 u_k \\ v_k
\end{pmatrix}
\end{equation}
and the linear term as
\begin{equation}
 - \sum_j \left[S^+_j u_j - \mathrm{i} S^-_j v_j\right] \,.
\end{equation}
We defined here the complex symmetric matrices $A_{jk}^\pm$ and the vectors $S^\pm_j$ by 
\begin{align}
A_{jk}^\pm &=  \frac{1}{2} \bigl( \widetilde{A}_{jk} \pm A_{jk} \bigr)   \,,
&
 S^\pm_j &= S_{j}[\theta] \pm S_{j}^*[\theta'{}] \mathrm{e}^{-\mathrm{i}\omega^e_j t}
\end{align}
Introducing the vector $\vek{w} = (u, v)^T$ where $u = (u_1,\dotsc ,u_{3N})$, $v = (v_1,\dotsc ,v_{3N})$, we can write the integral as
\begin{equation}
 \mathfrak I_{\alpha}(\lambda^g) = Z^2 \mathrm{e}^{\mathrm{i}\varphi} \mathrm{e}^{G^*(\theta')}\mathrm{e}^{G(\theta)} \int \frac{\mathrm{d}\vek{w}}{\uppi^{3N}}  \mathrm{e}^{ -\vek{w}^T . \vek{s}  - \vek{w}^T \Omega \vek{w} } \,,
\end{equation}
with
\begin{align}
 \Omega =& \begin{pmatrix}
      1 - \mathrm{A}^+ & \;\, -\mathrm{i} \mathrm{A}^- \\ -\mathrm{i} \mathrm{A}^- & 1 + \mathrm{A}^+
     \end{pmatrix} \,, &
 \vek{s} &= \begin{pmatrix}
            \phantom{-i}S^+ \\ -\mathrm{i} S^-
           \end{pmatrix} \,.
\end{align}
The result of the integral is given by 
\begin{equation}
  \int \frac{\mathrm{d}\vek{w}}{\uppi^{3N}}  \mathrm{e}^{ -\vek{w}^T . \vek{s}  - \vek{w}^T \Omega \vek{w} } = \sqrt{\frac{\uppi^{6N}}{\det \Omega}} \mathrm{e}^{\frac{1}{4} \vek{s}^T \Omega^{-1} \vek{s}} \,,
\end{equation}
and the integral in the $\alpha$'s, Eq.~\eqref{eq:integral_in_alpha}, reads
\begin{equation}
 \mathfrak I_{\alpha}(\lambda^g) = \frac{Z^2}{\sqrt{\det \Omega}} \mathrm{e}^{\mathrm{i}\varphi} \mathrm{e}^{G^*(\theta')}\mathrm{e}^{G(\theta)} \mathrm{e}^{\frac{1}{4} \vek{s}^T \Omega^{-1} \vek{s}} \,.
\end{equation}
For the demonstration of the convergence of the integral we refer the reader to Ref.~\cite{Baltrusch:2012}, where a proof has been reported.

\end{document}